\documentclass[pre,aps,preprint,amsmath,amssymb,floatfix,superscriptaddress]{revtex4-1}
\usepackage[utf8]{inputenc}
\usepackage{graphicx}
\usepackage{amsmath,amssymb}
\usepackage{array}
\newcolumntype{P}[1]{>{\centering\arraybackslash}p{#1}}

\usepackage[usenames]{color}
\newcommand{\red}[1]{\textcolor{black}{#1}}

\begin{document}

\title{Coupling volume-excluding compartment-based models of diffusion at different scales: Voronoi and pseudo-compartment approaches} 

\author{P. R. Taylor}
\email[]{paul.taylor@maths.ox.ac.uk}
\affiliation{Mathematical Institute, University of Oxford, Woodstock Road, Oxford, OX2 6GG, UK}

\author{R. E. Baker}
\affiliation{Mathematical Institute, University of Oxford, Woodstock Road, Oxford, OX2 6GG, UK}

\author{M. J. Simpson}
\affiliation{Mathematical Sciences, Queensland University of Technology, G.P.O. Box 2434, Brisbane 4001, Australia}

\author{C. A. Yates}
\affiliation{Department of Mathematical Sciences, University of Bath, Claverton Down, Bath, BA2 7AY, UK}

\date{\today}

\begin{abstract}
Numerous processes across both the physical and biological sciences are driven by diffusion. Partial differential equations (PDEs) are a popular tool for modelling such phenomena deterministically, but it is often necessary to use stochastic models to accurately capture the behaviour of a system, especially when the number of diffusing particles is low. The stochastic models we consider in this paper are `compartment-based': the domain is discretized into compartments, and particles can jump between these compartments. Volume-excluding effects (crowding) can be incorporated by blocking movement with some probability.

Recent work has established the connection between fine-grained models and coarse-grained models incorporating volume exclusion, but only for uniform lattices. In this paper we consider non-uniform, hybrid lattices that incorporate both fine- and coarse-grained regions, and present two different approaches to describing the interface of the regions. We test both techniques in a range of scenarios to establish their accuracy, benchmarking against fine-grained models, and show that the hybrid models developed in this paper can be significantly faster to simulate than the fine-grained models in certain situations, and are at least as fast otherwise. 
\end{abstract}

\maketitle

\section{Introduction}

Diffusion can be modelled using a multitude of mathematical techniques. Partial differential equations (PDEs) are a popular choice, but are inappropriate where stochastic effects are significant~\cite{Spill:2015:HRD}. Stochastic models of diffusion fall into two main categories: off-lattice models, where each particle's position lies on a continuum within the domain, and on-lattice, compartment-based models, where particles undergo a random walk on a lattice~\cite{Erban:2009:ABR}.

Volume exclusion (crowding) can be incorporated into compartment-based models by assigning a maximum particle capacity, $m$, to each compartment, where $m$ varies linearly with the compartment's size. Attempted moves into any given compartment are blocked with some probability. Of particular interest is the case where this probability scales linearly with the compartment's current occupancy, such that the blocking probability is equal to zero when empty and to unity when full to capacity, $m$~\cite{Painter:2003:VFQS,Anguige:2009:ODM,Khain:2014:SPO}. We describe such a model as `partially-excluding' or `coarse-grained' when $m>1$, and as `fully-excluding' or `fine-grained' when $m=1$. In one spatial dimension, the fully-excluding model is an example of single-file diffusion, a class of model of particular relevance to biological processes such as the diffusion of $\alpha\mathrm{TAT1}$ within microtubules~\cite{Farrell:2015:SFD}, the movement of flagellin in the formation of bacterial flagella~\cite{Stern:2013:DFF}, and the dispensing of proteins through in the nanochannels of drug delivery devices~\cite{Yang:2010:DCN}. 

In a recent paper~\cite{Taylor:2015:RTM}, we showed that the descriptions of a one-dimensional system given by fully-excluding and partially-excluding models can be reconciled. Specifically, we demonstrated that the mean and variance of particle numbers within each partially-excluding compartment of capacity $m$ can be matched with the mean and variance of the total number of particles across the $m$ corresponding contiguous fully-excluding compartments.

Our previous work considered uniform lattices, where every compartment was of the same size and capacity. For some models, such as those of real biological systems, it may be necessary to incorporate events occurring across a wide range of spatial and temporal scales~\cite{Walpole:2013:MCM}. This means that it may be desirable to, for example, simulate the diffusion of particles with precision in a small spatial region of interest, such as the area around receptors on a cell membrane~\cite{Klann:2012:HSG}, or around an ion channel~\cite{Flegg:2013:ICR}, while using less computationally intensive methods elsewhere on the domain. As a result, recent years have seen the development of multiple hybrid approaches for linking compartment-based models to off-lattice models of diffusion~\cite{Flegg:2012:TRM,Robinson:2014:ATM,Flegg:2015:CMM}, off-lattice models to PDE models of diffusion~\cite{Franz:2013:MRD}, and PDE models to compartment models of diffusion~\cite{Yates:2015:PCM,Ferm:2010:AAS,Moro:2004:HFP}. To our knowledge, however, there have been no attempts to develop a hybrid system interfacing volume-excluding compartment-based models at different scales.

In this paper, we consider how to interface a uniform region of partially-excluding compartments with another uniform region of fully-excluding compartments, and present two hybrid approaches capable of accurately simulating diffusion in such systems. Both approaches will be of value to researchers working on multi-scale systems, as they can speed up simulations while preserving precision where needed.

We begin by providing a short summary of our previous work on reconciling models of particle behaviour across spatial scales on uniformly partitioned lattices, before proceeding to consider hybrid models. The first hybrid method we present extends to diffusion on non-uniform lattices the previous results reconciling fully-excluding and partially-excluding systems. We use these results to  present one simple and elegant means of connecting two uniformly partitioned regions of differing compartment capacity. The second hybrid method we present defines a small area between the fine and coarse regions of the domain, where particle behaviour exhibits some characteristics of both regions. We term this area a pseudo-compartment, and the modelling framework a pseudo-compartment method, by analogy to a recent paper using similar techniques to couple a PDE model to a compartment-based model~\cite{Yates:2015:PCM}.

To determine the accuracy of these hybrid approaches, we apply each one to three scenarios, comparing the correspondence of the means and variances of particle numbers in the hybrid system to the corresponding moments in non-hybrid fine-grained and coarse-grained systems, as well as to PDE solutions when appropriate. The first scenario confirms that the hybrid systems are capable of maintaining a uniform steady state. The second compares their accuracy in a simple case of particle spreading from an initially inhomogeneous particle distribution. In the third scenario, we examine a morphogen gradient formation system, incorporating particle decay and influx of particles at the left-hand boundary $x=0$. Morphogen gradients are a common example of a multi-scale system in biology, as they incorporate both regions of low particle density where models with high spatial resolution are suitable, and regions of high particle density where such detailed modelling is computationally infeasible and unnecessary~\cite{Bollenback:2005:RFM,Tostevin:2007:FLP}. In all three cases, we observe that the mean and variance of particle numbers in each compartment of the hybrid models agree with those obtained from simulations run on uniform lattices.

Finally, we consider a simple multi-species scenario, where the simulated results of a partially-excluding model fail to match those of a fully-excluding model. We demonstrate that both hybrid systems are capable of matching the accuracy of the fully-excluding model in this scenario, while requiring only between a third and a seventh of the computational time to simulate. 

%In the case of the morphogen gradient, we also observe that both hybrid models are capable of providing significant time-savings compared to the fully-excluding model: the Voronoi and pseudo-compartment hybrid simulations finished in less than a quarter and less than a third, respectively, of the time required to simulate the fully-excluding model. Time savings in the other two test cases, although present, are much smaller, reflecting the fact that hybrid methods can only deliver significant efficiency when the majority of particle activity occurs within the coarse-grained region of the domain. 

\section{Random walks on uniform lattices}

We consider a one-dimensional lattice-based random walk on $x\in[0, L]$, where each motile particle has length $h$. The length of the domain, $L$, is chosen such that $L=Nh$, where $N\in\mathbb{N}$, so that the domain can contain at most $N$ particles. We impose a uniform lattice consisting of $K$ compartments onto this domain, where choices of $K$ are constrained by the requirement that $m=L/(Kh)$ should be a positive integer. As a consequence, $N=Km$, and $m$ describes the maximum number of particles that can reside in each of the $K$ compartments, or their `capacity'. Diffusion may then be modelled as a series of jumps as particles move between neighbouring compartments\red{; in the absence of volume exclusion, the jump rates between compartments scale inversely with the square of box lengths, i.e.}
\begin{equation}
 \red{\mathcal{T}_{j}^{\pm}=\dfrac{D}{m^2h^2}, \quad j=1,\ldots,K,}
\end{equation}
\red{where $D$ is the macroscale diffusivity of particles.}
\begin{figure}
\includegraphics[width=6cm]{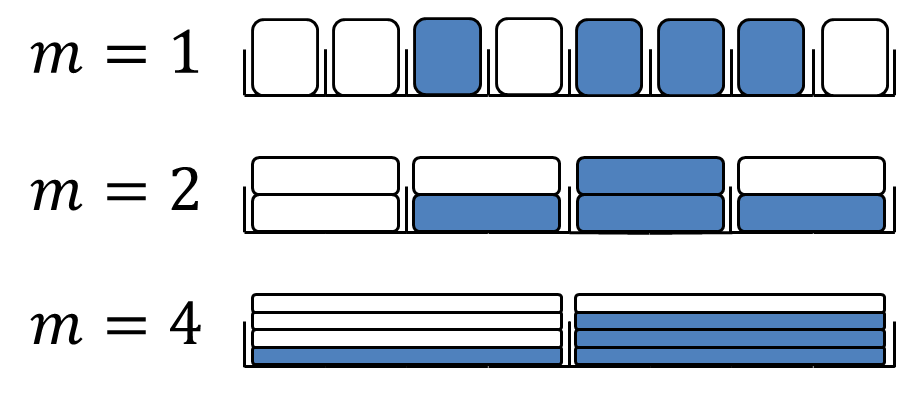}
\centering
\caption{Shaded cells represent particles, and white cells unused capacity, within uniform compartments of different capacities. As $m$ increases, the spatial resolution coarsens, but the descriptions given at these different scales can be reconciled.}
\label{fig:CroppedCoarseningLatticeDiagram}
\end{figure}
The spatial resolution of the model can be varied by changing the compartment capacity, $m$, as illustrated in Figure \ref{fig:CroppedCoarseningLatticeDiagram}. We are particularly interested in the fine-grained limiting case $m=1$ where each compartment contains at most one particle (`fully-excluding'). We shall consider the fully-excluding case to be `accurate', in the sense that no assumptions are required to determine each particle's position within its compartment. We assume that the particles considered in this paper do not interact except through volume-excluding effects.

\subsection{Volume exclusion}

A common approach taken in the literature is to define the jump rates between compartments as \red{the product of the non-excluding jump rate and a blocking probability:}
\begin{equation}
 T_j^{\pm}\red{=\mathcal{T}_j^{\pm}\left(1-\dfrac{n^{(m)}_{j\pm 1}}{m}\right)}=\dfrac{D}{m^2h^2}\left(1-\dfrac{n^{(m)}_{j\pm 1}}{m}\right), \quad j=1,\ldots,K,
\end{equation}
%where $D$ is the diffusivity of particles, 
\red{where} $n^{(m)}_{j}$ is the number of particles in compartment $j$ when each compartment has capacity $m$~\cite{Painter:2003:VFQS}. The term in brackets incorporates volume exclusion by causing attempted jumps into a compartment to fail with a probability that scales linearly with the compartment occupancy. Other forms for the blocking probability could be chosen, but the resulting particle behaviours would not be conserved across spatial scales~\cite{Taylor:2015:RTM}. Imposing zero-flux boundary conditions so that the number of particles in the system is constant, the evolution of mean particle numbers in compartment $1<j<K$ is given by
\begin{eqnarray}
 \label{eq:InitialMeanEvolutionEquationWithNonlinTerms}\dfrac{\mathrm{d}M_j}{\mathrm{d}t}=&&\red{\mathcal{T}_{j-1}^{+}}\left(M_{j-1}-\dfrac{1}{m}\langle n_{j-1}n_j\rangle\right)-\red{\mathcal{T}_{j}^{-}}\left(M_{j}-\dfrac{1}{m}\langle n_{j-1}n_j\rangle\right)\nonumber\\
 &-&\red{\mathcal{T}_{j}^{+}}\left(M_{j}-\dfrac{1}{m}\langle n_{j}n_{j+1}\rangle\right)+\red{\mathcal{T}_{j+1}^{-}}\left(M_{j+1}-\dfrac{1}{m}\langle n_{j}n_{j+1}\rangle\right).
 %=&&T^{+}_{j-1}M_{j-1}-\left(T^{-}_{j}+T^{+}_{j}\right)M_{j}+T^{-}_{j+1}M_{j+1}\nonumber\\
 %&+&\left(\dfrac{T^{-}_{j}}{m_{j-1}}-\dfrac{T^{+}_{j-1}}{m_{j}}\right)\langle n_{j-1}n_j\rangle+\left(\dfrac{T^{+}_{j}}{m_{j+1}}-\dfrac{T^{-}_{j+1}}{m_{j}}\right)\langle n_{j}n_{j+1}\rangle,\\
 \end{eqnarray}
where $\langle\cdot\rangle$ is used to denote expected values, and $M_j^{(m)}=\langle n_j^{(m)}\rangle$ denotes the mean number of particles in the $j^{th}$ compartment of capacity $m$. Writing out the transition rates explicitly, the evolution of mean particle numbers in each compartment is given by
\begin{eqnarray}
 \dfrac{\mathrm{d}M^{(m)}_1}{\mathrm{d}t}=&\dfrac{D}{m^2h^2}&\left(-M^{(m)}_{1}+M^{(m)}_{2}\right),\\
\label{eq:MeanMasterEquation} \dfrac{\mathrm{d}M^{(m)}_j}{\mathrm{d}t}=&\dfrac{D}{m^2h^2}&\left(M^{(m)}_{j-1}-2M^{(m)}_{j}+M^{(m)}_{j+1}\right), \quad j=2,\ldots,K-1,\\
 \dfrac{\mathrm{d}M^{(m)}_K}{\mathrm{d}t}=&\dfrac{D}{m^2h^2}&\left(M^{(m)}_{K-1}-M^{(m)}_{K}\right).
\end{eqnarray}
We refer to these as `mean master equations', and note their linearity: this is a consequence of our choice of blocking probability, and would not be the case had we chosen otherwise~\cite{Taylor:2015:RTM}. Non-linear terms will also occur when there is more than one species of particle present, or when there is directional bias in particle movement~\cite{Simpson:2009:MSE}. \red{The absence of non-linear terms in this macroscopic description does not imply that the individual particles are unconstrained by volume exclusion: a single tagged particle can display density dependent behaviour~\cite{Farrell:2015:SFD,Simpson:2009:MSE}.}

It is also possible to derive equations describing the evolution of the variance of $n_j^{(m)}$ (`variance master equations'):
\begin{eqnarray}
\label{eq:VarianceMasterEquation}
\dfrac{\mathrm{d}V_j^{(m)}}{\mathrm{d}t}
&=&
\frac{D}{m^2h^2}\left[\vphantom{\left(1-\frac{M_{j+1}^{(m)}}{m}\right)}2\left(\dfrac{m-1}{m}\right)V_{j,j-1}^{(m)}-4V_j^{(m)}+2\left(\dfrac{m-1}{m}\right)V_{j,j+1}^{(m)}
\vphantom{\left(1-\frac{M_{j}^{(m)}}{m}\right)}\right.
\nonumber\\
&&
+M_{j-1}^{(m)}\left(1-\frac{M_{j}^{(m)}}{m}\right)+M_j^{(m)}\left(1-\frac{M_{j-1}^{(m)}}{m}\right)\nonumber\\
&&
+\left.M_j^{(m)}\left(1-\frac{M_{j+1}^{(m)}}{m}\right)+M_{j+1}^{(m)}\left(1-\frac{M_{j}^{(m)}}{m}\right)\right],
\end{eqnarray}
for $2\leq{j}\leq{K-1}$, where $V^{(m)}_j=\langle (n^{(m)_j})^2\rangle-(M^{(m)}_j)^2$ is the variance of particle numbers in compartment $j$, and $V_{j,k}^{(m)}$ is the covariance of particle numbers in compartments $j$ and $k$:
\begin{eqnarray}
\label{eq:OffDiagCovarianceMasterEquation}
\dfrac{\mathrm{d}V_{j-1,j}^{(m)}}{\mathrm{d}t}
&=&
\frac{D}{m^2h^2}\left[\left(\dfrac{2}{m}-4\right)V_{j-1,j}^{(m)}+V_{j}^{(m)}+V_{j-1}^{(m)}+V_{j-2,j}^{(m)}+V_{j-1,j+1}^{(m)}
\vphantom{\left(1-\frac{M_{j}^{(m)}}{m}\right)}\right.
\nonumber\\
&&-\left.M_{j-1}^{(m)}\left(1-\dfrac{M_j^{(m)}}{m}\right)-M_j^{(m)}\left(1-\dfrac{M_{j-1}^{(m)}}{m}\right)\right];\\
\label{eq:DiffusionLikeCovarianceMasterEquation}
\dfrac{\mathrm{d}V_{j,k}^{(m)}}{\mathrm{d}t}
&=&\frac{D}{m^2h^2}\left[-4V_{j,k}^{(m)}+V_{j-1,k}^{(m)}+V_{j+1,k}^{(m)}+V_{j,k-1}^{(m)}+V_{j,k+1}^{(m)}\right] \qquad \text{for}\;\;\quad 1<j<k-1<K, \nonumber\\
&&\quad\quad\quad\quad\quad\quad\quad\quad\quad\quad\quad\quad\quad\quad\quad\quad\quad\quad\quad\quad\quad\quad\quad\quad \text{and}\quad 1<k+1<j<K.\nonumber
\end{eqnarray}
Similar expressions can be found for the variance and covariance terms involving the boundary compartments, $j=1,K$. \red{Note that, unlike the mean master equations, Eqs. (\ref{eq:VarianceMasterEquation}) and (\ref{eq:OffDiagCovarianceMasterEquation}) contain $m$ terms resulting from volume-excluding effects. As a result, mean particle numbers will evolve identically in volume-excluding and non-excluding models, but the variances and covariances of particle numbers will be different.}

When comparing a fully-excluding model ($m=1$) to a partially-excluding model ($m>1$), we seek to show that the mean and variance of particle numbers in each compartment is conserved. In order to compare the results of the two models, we first consider the means and variances of the total number of particles residing in a group of $m$ contiguous compartments with capacity one. We therefore define
\begin{equation}
S_{j}^{(m)}(t)=\sum\limits_{i=(j-1)m+1}^{jm}n_{i}^{(1)}(t),
\end{equation}
with $\mu_{j}^{(m)}(t)$ the mean of $S_{j}^{(m)}(t)$, and $v_{j}^{(m)}(t)$ its variance.

\subsection{Moving between scales}

\begin{figure}
\includegraphics[width=6cm]{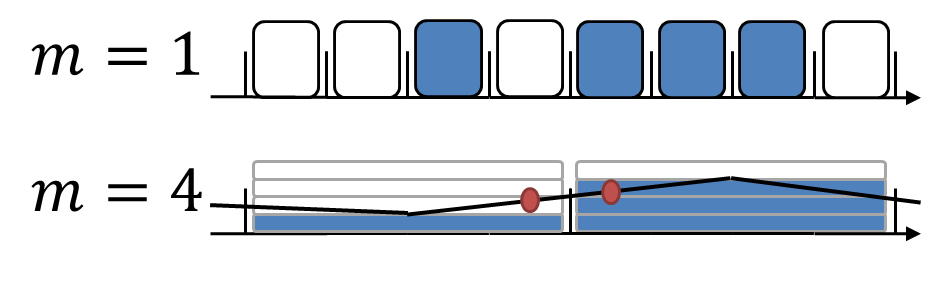}
\centering
\caption{Schematic of the interpolation process for $m=4$.}
\label{fig:interpolation}
\end{figure}

We wish to establish the relationship between: (i) $\mu_{j}^{(m)}(t)$ and $M_{j}^{(m)}(t)$; and (ii) $v_{j}^{(m)}(t)$ and $V_{j}^{(m)}(t)$. It is straightforward to show that, in both of these cases, the steady state values match. To find the relationship between $\mu_{j}^{(m)}(t)$ and $M_{j}^{(m)}(t)$ under transient conditions, we note
\begin{eqnarray}
\label{eq:SumOfExcludingMeanEquations}
\dfrac{\mathrm{d}\mu_{j}^{(m)}}{\mathrm{d}t}
&=&\sum\limits_{i=(j-1)m+1}^{jm}\dfrac{\mathrm{d}M_{i}^{(1)}}{\mathrm{d}t}\nonumber\\
&=&\dfrac{D}{h^2}\sum_{i=(j-1)m+1}^{jm}\left(M_{i-1}^{(1)}-2M_{i}^{(1)}+M_{i+1}^{(1)}\right)\nonumber\\
&=&\dfrac{D}{h^2}\left(M_{(j-1)m}^{(1)}-M_{(j-1)m+1}^{(1)}-M_{jm}^{(1)}+M_{jm+1}^{(1)}\right).
\end{eqnarray}
We compare this to the coarse-grained model, Eq.~\eqref{eq:MeanMasterEquation}, which we re-state here for convenience:
\begin{equation}
\nonumber
%\label{eq:PartiallyExcludingMeanEquationForComparisonPreExpansion}
\dfrac{\mathrm{d}M_{j}^{(m)}}{\mathrm{d}t}=\dfrac{D}{m^2h^{2}}\left(M_{j-1}^{(m)}-2M_{j}^{(m)}+M_{j+1}^{(m)}\right).
\end{equation}
Under the assumption that particles in the coarse ($m>1$) compartments are distributed, on average, following a linear interpolation between neighbouring coarse compartments. We therefore write
\begin{eqnarray}
\label{eq:FirstUnifInterpolationEquation}M_{jm}^{(1)}&=&\frac{1}{m}\left(\dfrac{1}{2}\dfrac{m+1}{m}\mu_{j}^{(m)}+\dfrac{1}{2}\dfrac{m-1}{m}\mu_{j+1}^{(m)}\right),\\
\label{eq:SecondUnifInterpolationEquation}M_{jm+1}^{(1)}&=&\frac{1}{m}\left(\dfrac{1}{2}\dfrac{m-1}{m}\mu_{j}^{(m)}+\dfrac{1}{2}\dfrac{m+1}{m}\mu_{j+1}^{(m)}\right),
\end{eqnarray}
as shown in Fig.~\ref{fig:interpolation}, with similar values for $M_{(j-1)m}^{(1)}$ and $M_{(j-1)m+1}^{(1)}$. As a result we have
\begin{equation}
\label{eq:AggregatedFineMeanEquation}\dfrac{\mathrm{d}\mu_{j}^{(m)}}{\mathrm{d}t}=\dfrac{D}{m^2h^{2}}\left(\mu_{j-1}^{(m)}-2\mu_{j}^{(m)}+\mu_{j+1}^{(m)}\right),
\end{equation}
and evolution of the mean particle numbers in the coarse-grained system matches that of the accurate model. Analogous reasoning can be applied to match the evolution of the variance of particle numbers. 

\section{Hybrid fully-/partially-excluding systems}

Having reviewed the reasoning used to match descriptions of uniform lattices at different spatial scales, we proceed to outline two hybrid models using non-uniform lattices. In both cases, we justify the correspondence of mean particle numbers in each compartment of the hybrid model to the mean particle numbers in the accurate, $m=1$ uniform lattice. 

\subsection{Voronoi method}

Jump rates between compartments can be derived from the assumption of underlying Brownian motion of particles using first-passage time arguments (the method is outlined in Appendix A)~\cite{Yates:2012:GMM}. Particles are assumed to begin from a fixed point within their current compartment, the `residence point', and are deemed to have jumped to a neighbouring compartment when they first reach that compartment's residence point. For a uniform lattice, these residence points are located at the centre of each compartment.

\begin{figure}
\includegraphics[width=10cm]{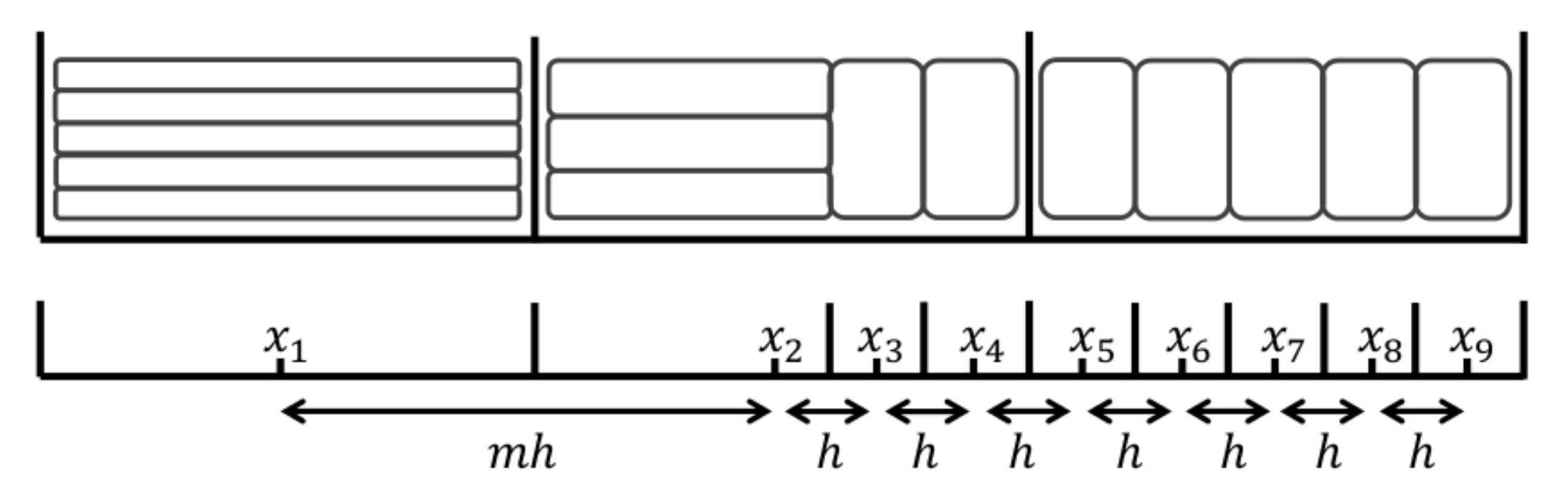}
\caption{The locations of a Voronoi lattice's residence points are chosen ($x_1,\dots,x_9$), and then compartment edges are positioned equidistant between neighbouring residence points. This particular arrangement demonstrates how a compartment of capacity $(m+1)/2$ can be used to interface regions where $m>1$ with regions where $m=1$. In this case, $\Delta x_2=mh$, and $\Delta x_j=h$ for $j=3,\dots,9$. When grouped, we would find $S_{1}^{(v)}(t)=n_{1}^{(v)}(t)$, $S_{2}^{(v)}(t)=\sum\limits_{i=2}^{4}n_{i}^{(v)}(t)$ and $S_{3}^{(v)}(t)=\sum\limits_{i=5}^{9}n_{i}^{(v)}(t)$.}
\label{fig:VoronoiHybridDiagram_Cropped}
\end{figure}

For a non-uniform lattice, we let the compartment residence points be at $[x_1,\dots,x_K]$ on the domain $[0,L]$, then define $\Delta x_j=x_j-x_{j-1}$, for $1<j\le K$. Compartment edges are then placed so as to be equidistant between neighbouring residence points, so that the $j^{th}$ compartment covers the interval $x\in[x_{j}-\Delta x_j/2, x_{j}+\Delta x_{j+1}/2]$ (an example is given in Figure \ref{fig:VoronoiHybridDiagram_Cropped}). Recall that the capacity of each compartment varies linearly with its length, and that the capacity of compartments on a uniform lattice is given by $m=L/(Kh)$, that is, the compartment length divided by particle length, $h$. Similarly, we obtain capacities $[m_1,\dots,m_K]$ for the $K$ Voronoi compartments by dividing their lengths by $h$, obtaining
\begin{eqnarray}
m_1&=&\dfrac{2x_1+\Delta x_2}{2h},\\
\label{eq:DefOfCompartmentCapacity}m_j&=&\dfrac{\Delta x_j+\Delta x_{j+1}}{2h},\;\;1<j<K,\\
m_K&=&\dfrac{2(L-x_K)+\Delta x_K}{2h}.
\end{eqnarray}
The positioning of residence points is restricted by a requirement that every compartment should have an integer-valued capacity. Inverting the conditional expected exit times obtained in Appendix A, and factoring in the relative probabilities of moving left or right, the following transition rates for a particle in the $j^{th}$ compartment\red{, in the absence of volume exclusion,} can be obtained~\cite{Redner:2001:FPP,Yates:2012:GMM}: 
\begin{eqnarray}
\label{eq:LeftVoronoiRate_WithoutVolumeExclusion}\mathcal{T}_j^{-}=&\dfrac{2D}{\Delta x_{j}(\Delta x_{j}+\Delta x_{j+1})}=\dfrac{D}{\Delta x_{j}m_jh},\\%&\left[1-\dfrac{n_{j-1}}{m_{j-1}}\right],\\
\label{eq:RightVoronoiRate_WithoutVolumeExclusion}\mathcal{T}_j^{+}=&\dfrac{2D}{\Delta x_{j+1}(\Delta x_{j}+\Delta x_{j+1})}=\dfrac{D}{\Delta x_{j+1}m_jh}.%&\left[1-\dfrac{n_{j+1}}{m_{j+1}}\right],
\end{eqnarray}
\red{Multiplying Eqs. (\ref{eq:LeftVoronoiRate_WithoutVolumeExclusion}) and (\ref{eq:RightVoronoiRate_WithoutVolumeExclusion}) by the jump blocking probability, $(1-n^{(v)}_{j\pm1}/m_{j\pm1})$, we arrive at the transition rates for a volume-excluding model:}
\begin{eqnarray}
 \label{eq:LeftVoronoiRate} \red{T_j^{-}}&\red{=}&\red{\mathcal{T}_{j}^{-}\left(1-\dfrac{n^{(v)}_{j-1}}{m_{j-1}}\right)=\dfrac{D}{\Delta x_{j}m_jh}\left(1-\dfrac{n^{(v)}_{j-1}}{m_{j-1}}\right),}\\%&\left[1-\dfrac{n_{j-1}}{m_{j-1}}\right],\\
 \label{eq:RightVoronoiRate}\red{T_j^{+}}&\red{=}&\red{\mathcal{T}_{j}^{+}\left(1-\dfrac{n^{(v)}_{j+1}}{m_{j+1}}\right)=\dfrac{D}{\Delta x_{j+1}m_jh}\left(1-\dfrac{n^{(v)}_{j+1}}{m_{j+1}}\right).}%&\left[1-\dfrac{n_{j+1}}{m_{j+1}}\right],
\end{eqnarray}
The following equation for the evolution of mean particle number can then be obtained
%\begin{eqnarray}
 %\label{eq:InitialVoronoiMeanEvolutionEquationWithNonlinTerms}\dfrac{\mathrm{d}M^{(v)}_j}{\mathrm{d}t}=&&T^{+}_{j-1}\left(M^{(v)}_{j-1}-\dfrac{1}{m_j}\langle %n^{(v)}_{j-1}n^{(v)}_j\rangle\right)-T^{-}_{j}\left(M^{(v)}_{j}-\dfrac{1}{m_{j-1}}\langle n^{(v)}_{j-1}n^{(v)}_j\rangle\right)\nonumber\\
 %&-&T^{+}_{j}\left(M^{(v)}_{j}-\dfrac{1}{m_{j+1}}\langle n^{(v)}_{j}n^{(v)}_{j+1}\rangle\right)+T^{-}_{j+1}\left(M^{(v)}_{j+1}-\dfrac{1}{m_{j}}\langle %n^{(v)}_{j}n^{(v)}_{j+1}\rangle\right),
 %\end{eqnarray}
 \begin{eqnarray}
 \label{eq:InitialVoronoiMeanEvolutionEquationWithNonlinTerms}\dfrac{\mathrm{d}M^{(v)}_j}{\mathrm{d}t}=&&\red{\mathcal{T}_{j-1}^{+}}\left(M^{(v)}_{j-1}-\dfrac{1}{m_j}\langle n^{(v)}_{j-1}n^{(v)}_j\rangle\right)-\red{\mathcal{T}_{j}^{-}}\left(M^{(v)}_{j}-\dfrac{1}{m_{j-1}}\langle n^{(v)}_{j-1}n^{(v)}_j\rangle\right)\\
 &-&\red{\mathcal{T}_{j}^{+}}\left(M^{(v)}_{j}-\dfrac{1}{m_{j+1}}\langle n^{(v)}_{j}n^{(v)}_{j+1}\rangle\right)+\red{\mathcal{T}_{j+1}^{-}}\left(M^{(v)}_{j+1}-\dfrac{1}{m_{j}}\langle n^{(v)}_{j}n^{(v)}_{j+1}\rangle\right),\nonumber
 \end{eqnarray}
 where we use $n^{(v)}_j$, $M^{(v)}_j$ and $V^{(v)}_j$ to denote the number, mean and variance, respectively, of particles in the $j^{th}$ Voronoi compartment. From the definitions of the transition rates given in Eqs. (\ref{eq:LeftVoronoiRate}) and (\ref{eq:RightVoronoiRate}) we note that $\red{\mathcal{T}}^{+}_{j-1}/m_j=\red{\mathcal{T}}^{-}_{j}/m_{j-1}$, and $\red{\mathcal{T}}^{+}_{j}/m_{j+1}=\red{\mathcal{T}}^{-}_{j+1}/m_{j}$, and hence Eq. (\ref{eq:InitialVoronoiMeanEvolutionEquationWithNonlinTerms}) simplifies to
 \begin{eqnarray}
 \dfrac{\mathrm{d}M^{(v)}_j}{\mathrm{d}t}%=&&T^{+}_{j-1}M_{j-1}-\left(T^{-}_{j}+T^{+}_{j}\right)M_{j}+T^{-}_{j+1}M_{j+1}\nonumber\\
 %&+&\left(\dfrac{Dh}{\Delta x_{j}m_jm_{j-1}}-\dfrac{Dh}{\Delta x_{j}m_{j-1}m_{j}}\right)\langle n_{j-1}n_j\rangle\nonumber\\
 %&+&\left(\dfrac{Dh}{\Delta x_{j+1}m_jm_{j+1}}-\dfrac{Dh}{\Delta x_{j+1}m_{j+1}m_{j}}\right)\langle n_{j}n_{j+1}\rangle,\\
 \label{eq:MeanVoronoiEquation}=&&\red{\mathcal{T}}^{+}_{j-1}M^{(v)}_{j-1}-\left(\red{\mathcal{T}}^{-}_{j}+\red{\mathcal{T}}^{+}_{j}\right)M^{(v)}_{j}+\red{\mathcal{T}}^{-}_{j+1}M^{(v)}_{j+1}.
\end{eqnarray}
In the same way, we can obtain equations for the evolution of variances,
\begin{eqnarray}\label{eq:GeneralisedVarianceEvolutionEquation}
 \dfrac{\mathrm{d}V^{(v)}_j}{\mathrm{d}t}&=&-2\left(\red{\mathcal{T}}^-_{j}+\red{\mathcal{T}}^+_{j}\right)V^{(v)}_j+2\red{\mathcal{T}}^+_{j-1}\left(1-\dfrac{1}{m_j}\right)V^{(v)}_{j-1,j}+2\red{\mathcal{T}}^-_{j+1}\left(1-\dfrac{1}{m_j}\right)V^{(v)}_{j,j+1}\nonumber\\
 &&+\red{\mathcal{T}}^+_{j-1}M^{(v)}_{j-1}\left(1-\dfrac{M^{(v)}_{j}}{m_{j}}\right)+\red{\mathcal{T}}^-_{j}M^{(v)}_{i}\left(1-\dfrac{M^{(v)}_{j-1}}{m_{j-1}}\right)\nonumber\\
 &&+\red{\mathcal{T}}^+_{j}M^{(v)}_{j}\left(1-\dfrac{M^{(v)}_{j+1}}{m_{j+1}}\right)+\red{\mathcal{T}}^-_{j+1}M^{(v)}_{j+1}\left(1-\dfrac{M^{(v)}_{j}}{m_{j}}\right).
\end{eqnarray}
The derivations of both Eq. (\ref{eq:InitialVoronoiMeanEvolutionEquationWithNonlinTerms}) and Eq. (\ref{eq:GeneralisedVarianceEvolutionEquation}) are outlined in Appendix \ref{appendix:VoronoiMasterEquationDerivation}. In the previous section, we summarised previous work matching the behaviours of $\mu_{j}^{(m)}(t)$ with $M_{j}^{(m)}(t)$, and of $v_{j}^{(m)}(t)$ with $V_{j}^{(m)}(t)$. We now wish to show that particle behaviour is similarly conserved between scales when using a Voronoi partitioned lattice.

Suppose that the Voronoi lattice consists of $p$ compartments of capacity $m>1$  at the left of the domain, followed by a single compartment of capacity $(m+1)/2$, and then the remainder of the domain is partitioned into fine compartments of capacity one (as illustrated in Figure \ref{fig:VoronoiHybridDiagram_Cropped}, for $p=1$ and $m=5$). To avoid non-integer compartment capacities, our choice of $m$ must be odd-valued. Recall that, to aggregate compartments on a fully-excluding uniform lattice, we wrote
\begin{equation}
S_{j}^{(m)}(t)=\sum\limits_{i=(j-1)m+1}^{jm}n_{i}^{(1)}(t).
\end{equation}
By analogy, we define $S_{j}^{(v)}(t)$, such that
%\begin{equation}
%S_{j}^{(v)}(t) =
%\left\{
%	\begin{array}{lll}
%		\quad\quad\quad\quad\quad\quad\; n^{(v)}_j(t)   &\mbox{if }& j < p+1 \\
%		\quad\quad\;\displaystyle\sum\limits_{i=p+1}^{p+(m+1)/2}n_{i}^{(v)}(t)  &\mbox{if }& j=p+1 \\
%		\displaystyle\sum\limits_{i=p+(m+3)/2+m(j-p-1)}^{p+(m+1)/2+m(j-p)}\hspace{-10 mm}n_{i}^{(v)}(t)  &\mbox{if }& j > p+1
%	\end{array}
%\right.
%\end{equation}
\begin{equation}
S_{j}^{(v)}(t) =
\left\{
	\begin{array}{lll}
		&&n^{(v)}_j(t)\quad\quad\quad\quad\quad\quad\mbox{if } j < p+1, \\
		&&\displaystyle\sum\limits_{i=p+1}^{p+(m+1)/2}\hspace{-4 mm}n_{i}^{(v)}(t)\quad\quad\quad\mbox{if } j=p+1, \\
		&&\displaystyle\sum\limits_{i=p+(m+3)/2+m(j-p-1)}^{p+(m+1)/2+m(j-p)}\hspace{-14 mm}n_{i}^{(v)}(t) \quad\!\!\mbox{if } j > p+1.
	\end{array}
\right.
\end{equation}
These groupings are illustrated in Figure \ref{fig:VoronoiHybridDiagram_Cropped} for $p=1$ and $m=5$. As in the previous section, we write $\mu_{j}^{(v)}(t)$ and $v_{j}^{(v)}(t)$ to denote the mean and variance of $S_{j}^{(v)}(t)$, and seek to show a connection between: (i) $\mu_{j}^{(v)}(t)$ and $\mu_{j}^{(m)}(t)$; and (ii) $v_{j}^{(v)}(t)$ and $v_{j}^{(m)}(t)$. We can do this by examining the equation for the evolution of $\mu^{(v)}_j$. For $j<p$, we write
\begin{eqnarray}
 \dfrac{\mathrm{d}\mu^{(v)}_j}{\mathrm{d}t}=\dfrac{\mathrm{d}M^{(v)}_j}{\mathrm{d}t}=\dfrac{D}{m^2h^{2}}\left(\mu_{j-1}^{(v)}-2\mu_{j}^{(v)}+\mu_{j+1}^{(v)}\right),
\end{eqnarray}
which clearly matches Eq. (\ref{eq:AggregatedFineMeanEquation}). For $j>p+1$, a similar match can be obtained using the same linear interpolation applied in the previous section.

When $j=p,p+1$ then we use Eqs. (\ref{eq:LeftVoronoiRate}) and (\ref{eq:RightVoronoiRate}) to obtain
\begin{eqnarray}
 \label{eq:LastCoarseVoronoiBoxMeans}
 \dfrac{\mathrm{d}\mu^{(v)}_p}{\mathrm{d}t}&=&\dfrac{D}{m^2h^{2}}M_{p-1}^{(v)}-2\dfrac{D}{m^2h^{2}}M_{p}^{(v)}+\dfrac{D}{mm_{p+1}h^2}M_{p+1}^{(v)},\\
  \dfrac{\mathrm{d}\mu^{(v)}_{p+1}}{\mathrm{d}t}&=&\sum\limits_{i=p+1}^{p+(m+1)/2}\dfrac{\mathrm{d}M^{(v)}_{i}}{\mathrm{d}t}(t)\nonumber\\
 \label{eq:FirstAggregatedVoronoiBoxMeans}&=&\dfrac{D}{m^2h^{2}}M_{p}^{(v)}-\dfrac{D}{mm_{p+1}h^2}M_{p+1}^{(v)}-\dfrac{D}{h^2}M^{(v)}_{p+(m+1)/2}+\dfrac{D}{h^2}M^{(v)}_{p+(m+3)/2}.
\end{eqnarray}
As before, we make the substitutions $M_{p-1}^{(v)}=\mu_{p-1}^{(v)}$ and $M_{p}^{(v)}=\mu_{p}^{(v)}$, and interpolate values of $M^{(v)}_{p+(m+1)/2}$ and $M^{(v)}_{p+(m+3)/2}$ from $\mu_{p+1}^{(v)}$ and $\mu_{p+2}^{(v)}$ using Eqs. (\ref{eq:FirstUnifInterpolationEquation}) and (\ref{eq:SecondUnifInterpolationEquation}). Finally, we observe that the residence point of the $(p+1)^{th}$ Voronoi box coincides with the centre of the $(p+1)^{th}$ aggregated box (as illustrated in Figure \ref{fig:VoronoiHybridDiagram_Cropped}), so the interpolated value for $M_{p+1}^{(v)}$ is simply
\begin{equation}
  M_{p+1}^{(v)}=\dfrac{m_{p+1}}{m}\mu_{p+1}^{(v)}.                                                                                                                                                                                                                                                                                                                                                                                                                                                                                                                                                                                                                                                                                                                      \end{equation}
Substituting this into Eqs. (\ref{eq:LastCoarseVoronoiBoxMeans}) and (\ref{eq:FirstAggregatedVoronoiBoxMeans}), we arrive at
\begin{eqnarray}
 \dfrac{\mathrm{d}\mu^{(v)}_p}{\mathrm{d}t}&=&\dfrac{D}{m^2h^{2}}\left(\mu_{p-1}^{(v)}-2\mu_{p}^{(v)}+\mu_{p+1}^{(v)}\right),\\
  \dfrac{\mathrm{d}\mu^{(v)}_{p+1}}{\mathrm{d}t}&=&\dfrac{D}{m^2h^{2}}\left(M_{p}^{(v)}-2\mu^{(v)}_{p+1}+\mu^{(v)}_{p+2}\right),
\end{eqnarray}
so we can expect the values of $\mu^{(v)}$ to match those of $\mu^{(m)}$. Although we do not show them here, analogous treatments can be applied to match $v^{(v)}$ with $v^{(m)}$.

Voronoi partitioned lattices provide one method of interfacing two regions partitioned by uniform lattices of differing compartment capacities, but only for certain compartment capacity values. It is not possible, for example, to interface a region where $m=1$ with a region where $m=2$, as any intermediate sized compartment would have non-integer valued capacity. In such situations, it is useful to have another hybrid method capable of interfacing the regions.

\subsection{Pseudo-compartment method}

Pseudo-compartment methods have previously been used to interface compartment-based models with PDE models. In such models, the domain consists of a region where diffusion is modelled using a discrete model, another region in which diffusion is modelled using a continuum PDE model, and a pseudo-compartment that combines both models. Specifically, although particle concentration in the pseudo-compartment is modelled in a spatially continuous manner using PDEs, discrete jump events between the pseudo-compartment and the neighbouring compartment-based region can take place, adjusting the particle concentration profile appropriately~\cite{Yates:2015:PCM}. In this section, we discuss using similar ideas to interface one region of the domain, partitioned by a coarse lattice with compartment capacity $m>1$, with another region, partitioned with a fine lattice where $m=1$. 

\begin{figure}
\includegraphics[width=10cm]{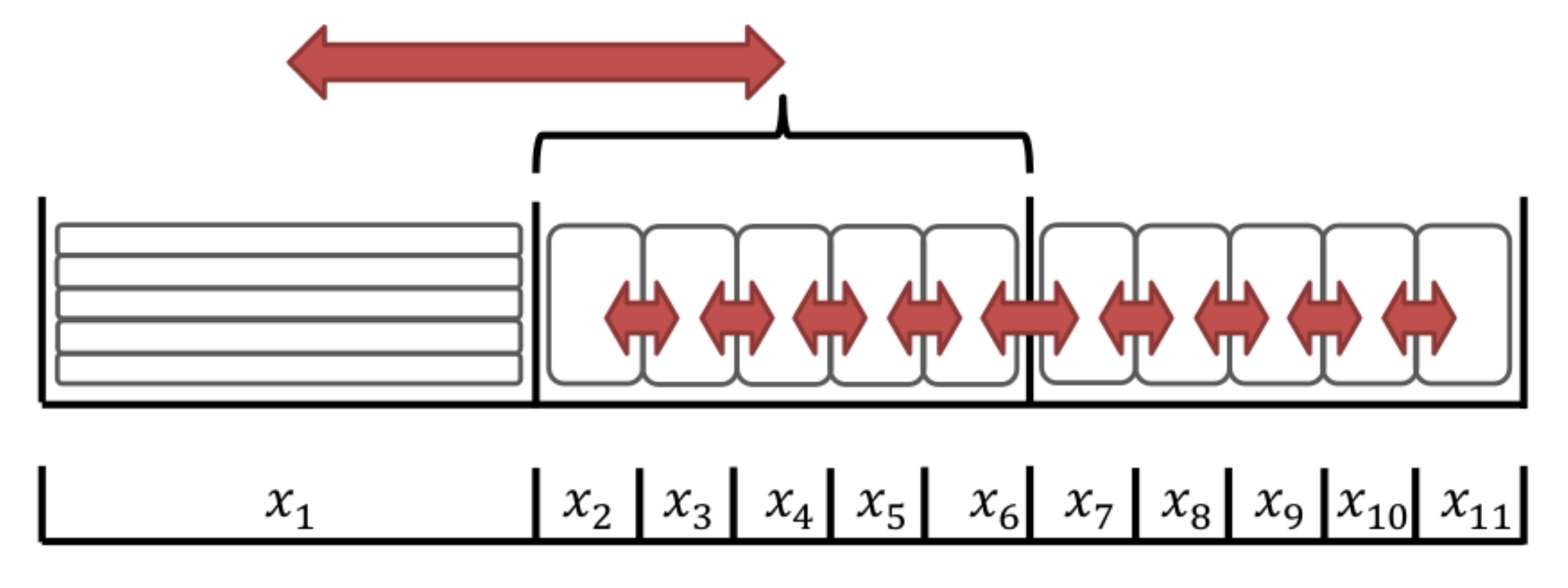}
\caption{Pseudo-compartment method, with $m=5$ and $p=1$. The compartments of index $2$ through to $6$ form the pseudo-compartment. Particles in these compartments may attempt to jump to neighbouring compartments with rate $D/h^2$, or into compartment $1$ with propensity $D/m^2h^2$, subject to the standard blocking probability arising from volume exclusion. Particles in compartment $1$ may jump into any unoccupied compartment with index $2$ through to $6$ with propensity $D/mh^2$. When grouped, we have $S_{1}^{(c)}(t)=n_{1}^{(c)}(t)$, $S_{2}^{(c)}(t)=\sum\limits_{i=2}^{6}n_{i}^{(c)}(t)$ and $S_{3}^{(c)}(t)=\sum\limits_{i=7}^{11}n_{i}^{(c)}(t)$.}
\label{fig:PseudoCompartmentDiagram_Cropped}
\end{figure}

We assume, without loss of generality, that the coarse region is to the left hand side of the domain, and that it consists of $p$ compartments of capacity $m>1$. The first $m$ fully-excluding compartments, i.e. those of index $p+1$ through to $p+m$, are then said to collectively comprise the pseudo-compartment (this is illustrated for $p=1$ and $m=5$ in Figure \ref{fig:PseudoCompartmentDiagram_Cropped}). Particles in compartments $1$ through to $p$ may attempt to jump to neighbouring compartments in this range with rate $D/m^2h^2$, while particles in compartments $p+1$ through to $K$ may attempt to jump to neighbouring compartments in this range with rate $D/h^2$. In both cases, these jumps fail with blocking probability,
\begin{equation}
 \left(1-\dfrac{n_{j}^{(c)}}{m_j}\right),
\end{equation}
where $j$ is the index of the destination compartment, and we use $n_j^{(c)}$, and later $M_p^{(c)}$, to denote, respectively the number of particles and the mean number of particles in compartment $j$ of the lattice.

In addition, it is possible for a particle in compartment $p$ to move into the pseudo-compartment with jump propensity $D/m^2h^2$. When this happens, one of the compartments that comprise the pseudo-compartment is selected uniformly at random, and if it is unoccupied a particle jumps into it from compartment $p$ (if it is occupied then the jump is terminated). Equivalently, volume-excluding effects can be incorporated by multiplying the jump propensity by the fraction of compartments in the pseudo-compartment currently unoccupied, and selecting a destination from among the unoccupied compartments uniformly at random when a jump occurs.

Similarly, a particle residing within any part of the pseudo-compartment may jump into compartment $p$, with propensity
\begin{equation}
 \dfrac{D}{m^2h^2}\left(1-\dfrac{n_p^{(c)}}{m}\right).
\end{equation}
All of the possible particle jumps for an example lattice are illustrated by arrows in Figure \ref{fig:PseudoCompartmentDiagram_Cropped}.
%This is in addition to the standard jump propensities for movement to adjacent compartments where $m=1$ with the standard rate,
%\begin{equation}
% \dfrac{D}{h^2}\left(1-n_p^{(c)}\right)
%\end{equation}
%where $j$ denotes the index of the destination compartment.
Suppose that the $p^{th}$ compartment is the last partially-excluding compartment, so that all compartments of index greater than $p$ will be fully-excluding. Within the coarse region, we write the usual equations for the evolution of mean particle number
\begin{eqnarray}
 \dfrac{\mathrm{d}M^{(c)}_1}{\mathrm{d}t}&=&\dfrac{D}{m^2h^2}\left(-M^{(c)}_1+M^{(c)}_2\right),\\
 \dfrac{\mathrm{d}M^{(c)}_i}{\mathrm{d}t}&=&\dfrac{D}{m^2h^2}\left(M^{(c)}_{i-1}-2M^{(c)}_{i}+M^{(c)}_{i+1}\right),\;\;1<i<p.
\end{eqnarray}
The mean master equation for compartment $p$, the final partially-excluding compartment, is modified to account for the incoming particles it can receive from all fully-excluding compartments within the pseudo-compartment:
\begin{equation}
 \label{eq:EquationForPseudoBoxP}\dfrac{\mathrm{d}M^{(c)}_p}{\mathrm{d}t}=\dfrac{D}{m^2h^2}\left(M^{(c)}_{p-1}-2M^{(c)}_{p}+\sum\limits_{j=p+1}^{p+m}M^{(c)}_{j}\right).
\end{equation}
Similarly, the equations for mean particle number in the $m$ fully-excluding compartments comprising the pseudo-compartment region must also account for particles moving to, and arriving from, the partially-excluding region, as well as for movement over the fully-excluding lattice,
\begin{eqnarray} 
 \label{eq:EquationForPseudoBoxPPlusOne}\dfrac{\mathrm{d}M^{(c)}_{p+1}}{\mathrm{d}t}&=&\dfrac{D}{h^2}\left(\dfrac{1}{m}\dfrac{1}{m^2}M^{(c)}_{p}-\left[1+\dfrac{1}{m^2}\right]M^{(c)}_{p+1}+M^{(c)}_{p+2}\right),\\
 \label{eq:EquationForPseudoBoxPPlusSome}\dfrac{\mathrm{d}M^{(c)}_{i}}{\mathrm{d}t}&=&\dfrac{D}{h^2}\left(\dfrac{1}{m}\dfrac{1}{m^2}M^{(c)}_{p}+M^{(c)}_{i-1}-\left[2+\dfrac{1}{m^2}\right]M^{(c)}_{i}+M^{(c)}_{i+1}\right),\;\;p+1<i\le p+m,
\end{eqnarray}
and, finally, beyond the pseudo-compartment, the evolution of mean particle number is described using the standard mean master equations with $m=1$,
\begin{eqnarray} 
 \dfrac{\mathrm{d}M^{(c)}_i}{\mathrm{d}t}&=&\dfrac{D}{h^2}\left(M^{(c)}_{i-1}-2M^{(c)}_{i}+M^{(c)}_{i+1}\right),\;\;p+m<i<K,\\
 \dfrac{\mathrm{d}M^{(c)}_K}{\mathrm{d}t}&=&\dfrac{D}{h^2}\left(M^{(c)}_{K-1}-M^{(c)}_{K}\right).
\end{eqnarray}
To compare the description of particle diffusion given by the pseudo-compartment model to the description given by the fully-excluding model where $m=1$ we must aggregate the compartments, writing
\begin{equation}
S_{j}^{(c)}(t) =
\left\{
	\begin{array}{lll}
		&&n^{(c)}_j(t) \quad\quad\quad\quad\quad \mbox{if } j < p+1, \\
		&&\displaystyle\sum\limits_{i=p+1}^{p+m}n_{i}^{(c)}(t)\quad\quad\quad\!\mbox{if } j=p+1, \\
		&&\displaystyle\sum\limits_{i=p+1+m(j-p-1)}^{p+m(j-p)}\hspace{-8 mm}n_{i}^{(c)}(t)\quad\!\mbox{if } j > p+1.
	\end{array}
\right.
\end{equation}
%\begin{equation}
%S_{j}^{(c)}(t) =
%\left\{
%	\begin{array}{lll}
%		&n^{(c)}_j(t)  & \mbox{if } j < p+1 \\
%		\quad\quad\displaystyle\sum\limits_{i=p+1}^{p+m}&n_{i}^{(c)}(t) & \mbox{if } j=p+1 \\
%		\displaystyle\sum\limits_{i=p+1+m(j-p-1)}^{p+m(j-p)}&n_{i}^{(c)}(t) & \mbox{if } j > p+1
%	\end{array}
%\right.
%\end{equation}
As before, we write $\mu_{j}^{(c)}(t)=\langle S_{j}^{(c)}(t)\rangle$, and seek to match $\mu_{j}^{(c)}(t)$ with $\mu_{j}^{(m)}(t)$. This is trivial for $j<p$ and for $j>p+1$, as these are sections of uniformly partitioned compartments where $m>1$ and $m=1$, respectively, and so the previously obtained results apply. When $j=p$, we use Eq. (\ref{eq:EquationForPseudoBoxP}) to write
\begin{eqnarray}
 \dfrac{\mathrm{d}\mu^{(c)}_{p}}{\mathrm{d}t}=\dfrac{\mathrm{d}M^{(c)}_p}{\mathrm{d}t}&=&\dfrac{D}{m^2h^2}\left(M^{(c)}_{p-1}-2M^{(c)}_{p}+\sum\limits_{j=p+1}^{p+m}M^{(c)}_{j}\right),\nonumber\\
 &=&\dfrac{D}{m^2h^2}\left(\mu^{(c)}_{p-1}-2\mu^{(c)}_{p}+\mu^{(c)}_{p+1}\right),
\end{eqnarray}
which is the expected form needed to match $\mu^{(c)}_{p}$ with $\mu^{(m)}_{p}$. Similarly, for $j=1$ we use Eqs. (\ref{eq:EquationForPseudoBoxPPlusOne}) and (\ref{eq:EquationForPseudoBoxPPlusSome}) to write
\begin{eqnarray}
 \dfrac{\mathrm{d}\mu^{(c)}_{p+1}}{\mathrm{d}t}=\sum\limits_{j=p+1}^{p+m}\dfrac{\mathrm{d}M^{(c)}_j}{\mathrm{d}t}&=&\dfrac{D}{m^2h^2}M^{(c)}_p-\dfrac{D}{m^2h^2}\sum\limits_{j=p+1}^{p+m}M^{(c)}_{j}-\dfrac{D}{h^2}M^{(c)}_{p+m}+\dfrac{D}{h^2}M^{(c)}_{p+m+1}\nonumber\\
 &=&\dfrac{D}{m^2h^2}\mu^{(c)}_p-\dfrac{D}{m^2h^2}\mu^{(c)}_{j}-\dfrac{D}{h^2}M^{(c)}_{p+m}+\dfrac{D}{h^2}M^{(c)}_{p+m+1}.
\end{eqnarray}
Interpolating values of $M^{(c)}_{p+m}$ and $M^{(c)}_{p+m+1}$ using Eqs.  (\ref{eq:FirstUnifInterpolationEquation}) and (\ref{eq:SecondUnifInterpolationEquation}), we finally arrive at
\begin{equation}
 \dfrac{\mathrm{d}\mu^{(c)}_{p+1}}{\mathrm{d}t}=\dfrac{D}{m^2h^2}\left(\mu^{(c)}_{p}-2\mu^{(c)}_{p+1}+\mu^{(c)}_{p+2}\right),
\end{equation}
so we expect the values of $\mu^{(c)}_{j}$ to match those of $\mu^{(m)}_{j}$.

Having presented two hybrid methods for interfacing partially-excluding with fully-excluding regions, we now apply them both to a number of test systems to check their correspondence to the fully-excluding model where $m=1$.

\section{Numerical investigations}
\label{section:NumericalInvestigations}
In this section we consider three single-species test cases: maintaining a uniform steady state in a purely diffusive system; particle redistribution from a non-uniform initial state; and a morphogen gradient. For each case we compare four different models:
\begin{enumerate}
 \item Uniform fully excluding ($m=1$): 105 compartments of capacity one and length 0.2.
 \item Uniform partially excluding ($m=7$): 15 compartments of capacity seven and length 1.4.
 \item Hybrid Voronoi: five compartments of capacity seven and length 1.4, one box of capacity four and length 0.8, and 66 compartments of capacity one and length 0.2.
 \item Hybrid pseudo-compartment: five compartments of capacity seven and length 1.4, and 70 of capacity one and length 0.2. The first seven fully-excluding compartments form the pseudo-compartment.
\end{enumerate}
In each case we use these four models to obtain, respectively, numerical values for $\mu_j^{(m)}$ and $v_j^{(m)}$, for $M_j^{(m)}$ and $V_j^{(m)}$, for $\mu_j^{(v)}$ and $v_j^{(v)}$, and for $\mu_j^{(p)}$ and $v_j^{(p)}$, aggregating compartments from the finer grids into regions of capacity $m=7$. In each case we set $h=0.2$, $L=21$, and hence $N=105$. We choose diffusion constant $D=2$ for all test cases, and perform $50,000$ realisations of each case over the time interval $t\in[0,25]$. 

Since we consider the fully-excluding model as `accurate', in the sense that it specifies each particle's location precisely rather than making assumptions about its position within a larger compartment, a further $50,000$ realisations of the fully-excluding model were generated to provide an independent estimate for $\mu_j^{(m)}$ and $v_j^{(m)}$, and these values were used as our comparison data set. For the first two single-species test cases, it would also be possible to generate a comparison data set by evaluating the master equations deterministically, but this would not be possible for the morphogen gradient test (the addition of particles to the first compartment with available capacity leads to a non-closed system of equations). In the interests of consistency, we have therefore generated all data sets by stochastic simulation, rather than using a mixture of approaches.

The agreement of the four models with the comparison data set was then quantified using the Histogram Distance Error (HDE)~\cite{Cao:2006:ALA}:
\begin{equation}
 \mathrm{HDE}=\frac{1}{2}\sum\limits_{k=1}^{K}|s_k-c_k|,
\end{equation}
where $s_k$ is the normalised value of the $k^\text{th}$ aggregated compartment of the model being considered, and $c_k$ is the normalised value of the $k^\text{th}$ aggregated compartment of the comparison data set, such that
\begin{equation}
 \sum\limits_{k=1}^{K}s_k=\sum\limits_{k=1}^{K}c_k=1.
\end{equation}
The HDE therefore returns values between zero and one, where zero corresponds to two identical data sets and one represents two completely distinct data sets. We note that the HDE is an example of an $L^{1}$-norm, but that qualitatively similar results are observed when using $L^{2}$-norms. Our conclusions are not affected by the choice of using an $L^{1}$ or an $L^{2}$-norm.

To generate realisations we used an algorithm based on Gillespie's Direct Method~\cite{Gillespie:1977:ESS}, as described in Appendix \ref{appendix:SimulationAlgorithm}. In each of the three test cases, we also record the time taken to simulate 50,000 realisations of each model. All simulations were performed using Matlab code, parallelised using \textsf{parfor}, on a four-core desktop computer using an AMD Phenom\textsuperscript{TM} II X4 925 Processor.

\subsection{Test Case 1: Maintaining a spatially uniform steady state}

In this simplest test, for each realisation we initialise 15 uniformly distributed particles over the lattice, let them diffuse freely, then record their final positions to ensure they remain uniformly distributed. This is the most basic test, and is performed to confirm that the hybrid interfaces do not interfere with a homogeneous particle distribution.

\begin{figure}
\includegraphics[width=18cm]{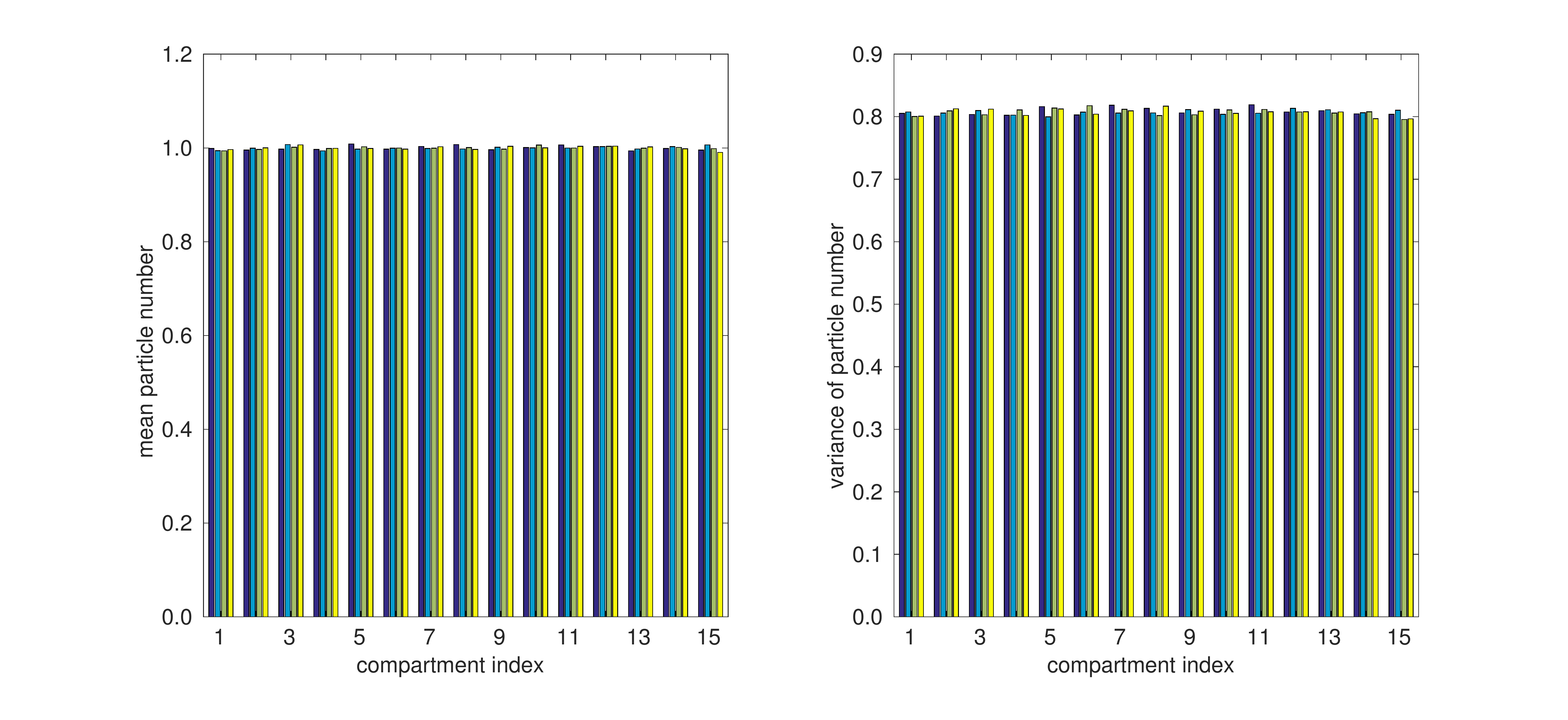}
\caption{Results from 50,000 repeats of the uniform steady state simulation at time $t=25$. Fifteen particles were uniformly distributed at time $t=0$ and then diffused until time $t=25$. From left to right, each set of four bars shows results from the fully-excluding (dark blue), partially-excluding (light blue), Voronoi (green) and pseudo-compartment (yellow) models.}
\label{fig:USS_MeanAndVar}
\end{figure}

The mean and variance of particle numbers in each aggregated box at $t=25$ are plotted in Figure \ref{fig:USS_MeanAndVar}, showing good agreement between all four models. This agreement is quantified in Table \ref{table:UniformSteadyState}, where we observe that in each model, mean values deviate from the comparison data by less than $0.25\%$, and variance values deviate by less than $0.5\%$. Although we do not present the details here, it is also possible to demonstrate the steady state agreement of all four models by analysing their mean and variance master equations.

We find that the hybrid models are faster to simulate than the fully-excluding model. However, the decrease in computational time is modest, and we anticipate that significant computational savings will only be achieved when the majority of particle motion is concentrated in the partially-excluding region. The case of maintaining a spatially uniform steady state is used in this paper solely to check the accuracy of the hybrid models, and hybrid modelling will not be helpful for such systems, unless the region partitioned with fully-excluding compartments is small relative to the domain as a whole.
%Some time-savings for the hybrid models are observed compared to the uniform, fully-excluding model, though as expected they are relatively small (we only anticipate significant time-savings when the majority of particle activity is concentrated in the partially-excluding region to the left of the domain).

\begin{center}
\begin{table}
\caption{Details of uniform distribution simulation.}
    \begin{tabular}{ | l || P{20mm} | P{25mm} | P{35mm} | P{40mm} |}
    \hline
    Model & Mean HDE  & Variance HDE & Time to simulate & Acceleration relative \\
	  & at $t=25$	& at $t=25$ && to fully-excluding	\\ \hline \hline
    Fully-excluding 	& 0.0023 & 0.0046 & 28741 seconds 	& - 			\\ \hline %7 hours, 59 minutes, 1 seconds  \\ \hline
    Partially-excluding & 0.0022 & 0.0036 & 582 seconds		& 49.4 times faster 	\\ \hline %9 minutes, 42 seconds  \\ \hline
    Voronoi 		& 0.0020 & 0.0035 & 19738 seconds 	& 1.5 times faster 	\\ \hline %5 hours, 28 minutes, 58 seconds  \\ \hline
    Pseudo-compartment & 0.0017 & 0.0033 & 26361 seconds 	& 1.1 times faster 	\\ \hline %7 hours, 19 minutes, 21 seconds  \\ \hline
    \end{tabular}
    \label{table:UniformSteadyState}
\end{table}
\end{center}

\subsection{Test Case 2: Particle redistribution}

The second test case examines the ability of our hybrid models to accommodate a net particle flux over the interface. Our initial state consists of thirty-five particles collected at the left of the domain. For the uniform fully-excluding model, this means that the first thirty-five compartments are occupied. For all the other models, the first five compartments, each of capacity seven, are full to capacity.

\begin{figure}
\includegraphics[width=16cm]{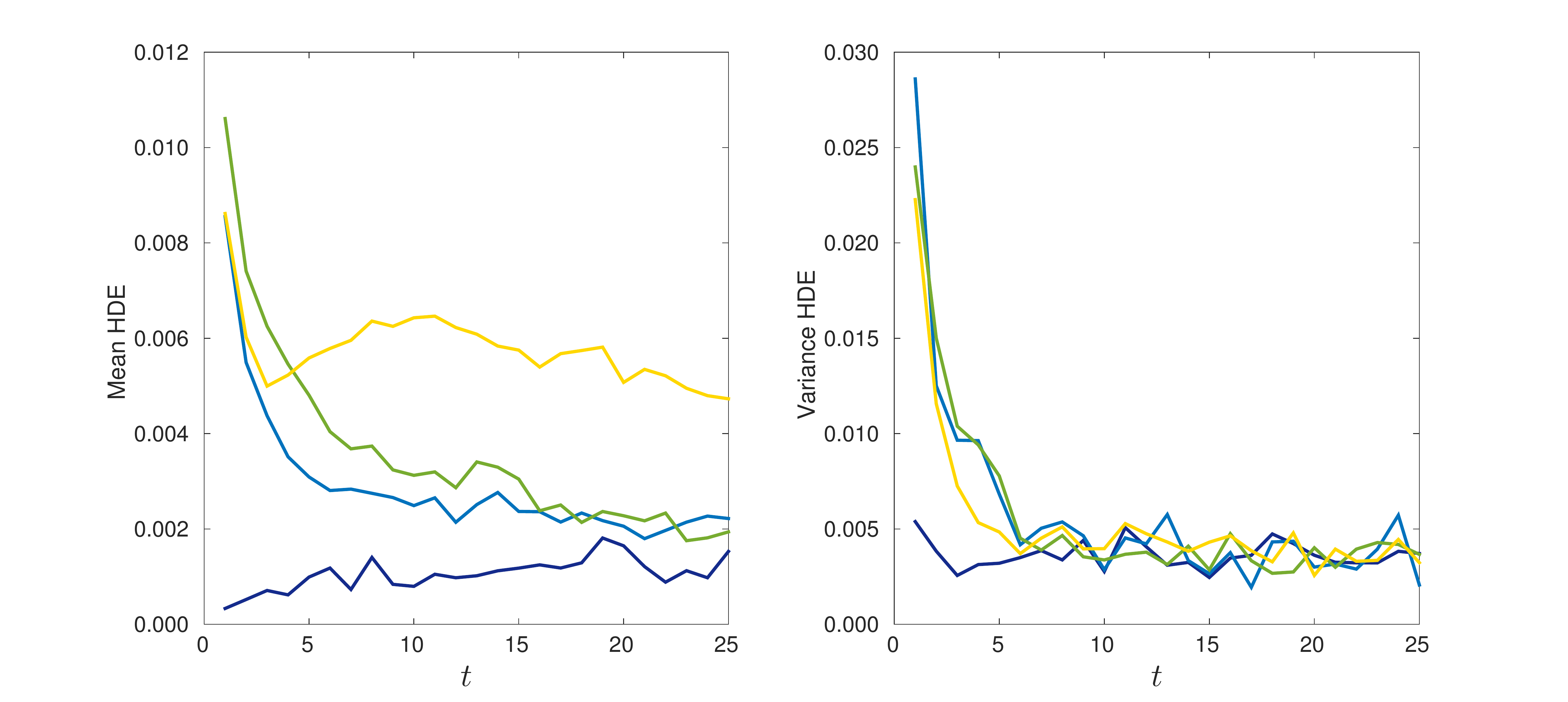}
\caption{HDE values for the mean and variance of the particle redistribution simulation results at discrete times $t=1,2,\dots,25$. The four data sets in each figure correspond to the fully-excluding (dark blue), partially-excluding (light blue), Voronoi (green) and pseudo-compartment (yellow) models.}
\label{fig:FrontSpreading_HDE_Comparison}
\end{figure}

We compare the agreement of mean and variance values at $t=1,2,3\dots,25$ in Figure \ref{fig:FrontSpreading_HDE_Comparison}. As anticipated, all four models perform well, although the Voronoi model performs somewhat better than the pseudo-compartment model. A possible explanation of this is that, with particle concentrations in the pseudo-compartment higher to the left and lower to the right, particles jumping into the pseudo-compartment from compartment $p$ will disproportionately arrive in the more numerous vacant compartments to the right of the pseudo-compartment, slightly lengthening the average jump length. We note though that, even in this case of strongly asymmetric flux, the HDE remains below 0.008 throughout the simulation.

\begin{figure}
\includegraphics[width=18cm]{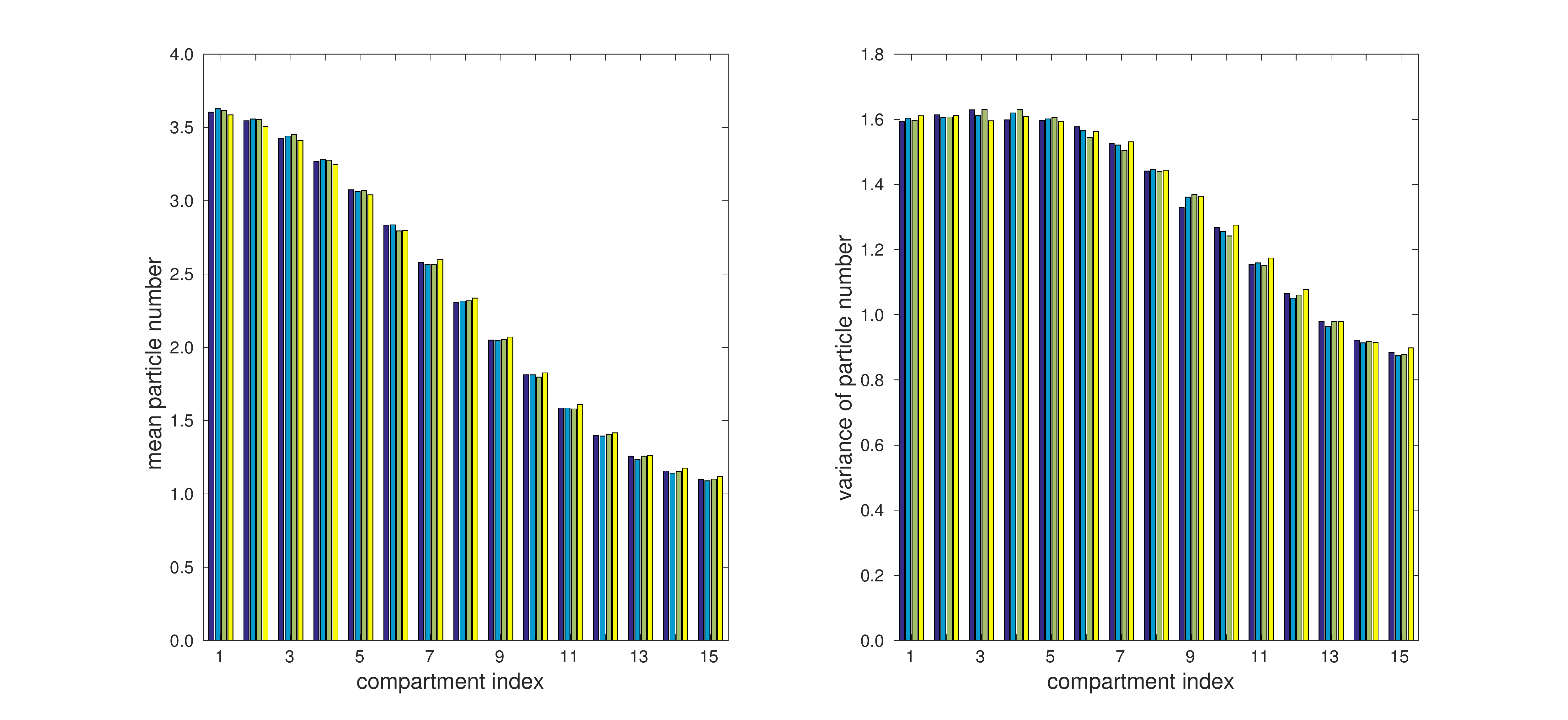}
\caption{Results from 50,000 repeats of the particle redistribution simulation at time $t=25$. Thirty-five particles were initialised at the left of the domain at time $t=0$. From left to right, each set of four bars shows results from the fully-excluding (dark blue), partially-excluding (light blue), Voronoi (green) and pseudo-compartment (yellow) models.}
\label{fig:FrontSpreading_MeanAndVar_t25}
\end{figure}

We also plot the means and variances of particle numbers at $t=25$ in Figure \ref{fig:FrontSpreading_MeanAndVar_t25}, and for consistency with the other test cases list the HDE values at time $t=25$ in Table \ref{table:FrontSpreading}. We also record the time taken to simulate each model in Table \ref{table:FrontSpreading}, noting again that the hybrid models run faster than the fully-excluding model, with the Voronoi model taking only half as much time to simulate. As with the spatially uniform steady state considered in the previous test, however, we would only expect significant time savings when the fully-excluding region of the domain is relatively small; this case is presented primarily as a test of accuracy rather than to demonstrate significantly accelerated computation.
\begin{center}
\begin{table}
\caption{Details of particle redistribution simulation.}
    \begin{tabular}{ | l || P{20mm} | P{25mm} | P{35mm} | P{40mm} |}
    \hline
    Model & Mean HDE  & Variance HDE & Time to simulate & Acceleration relative \\
	  & at $t=25$	& at $t=25$ &	& to fully-excluding \\ \hline \hline
    Fully-excluding 	& 0.0015 & 0.0037 & 40775 seconds & -  			\\ \hline%11 hours, 19 minutes, 35 seconds  \\ \hline
    Partially-excluding & 0.0022 & 0.0021 & 752 seconds   & 54.2 times faster	\\ \hline%12 minutes, 32 seconds  \\ \hline
    Voronoi 		& 0.0019 & 0.0037 & 20884 seconds & 2.0 times faster 	\\ \hline%5 hours, 48 minutes, 4 seconds  \\ \hline
    Pseudo-compartment & 0.0047 & 0.0032 & 31330 seconds & 1.3 times faster 	\\ \hline%8 hours, 42 minutes, 10 seconds  \\ \hline
    \end{tabular}
    \label{table:FrontSpreading}
\end{table}
\end{center}

\subsection{Test Case 3: Morphogen gradient}
\label{subsec:MorphogenGradTest}
In the final single-species test case, a simple morphogen gradient system is simulated. Starting from an empty initial state, particles enter the domain at the left-hand boundary with rate $r_1$, while a zero-flux boundary condition is imposed at the right-hand boundary. As well as moving diffusively, particles decay with rate $r_2$. In the absence of volume exclusion, a flux boundary condition can be implemented by adding or removing particles in the closest compartment to the boundary at a specified rate~\cite{Taylor:2015:NLJ}. When volume exclusion is incorporated, however, it will sometimes be the case that the closest compartment to the boundary is already filled to capacity, in which case no more particles may be added to it. When this occurred in simulations, we added particles to the first compartment from the boundary with available capacity.

\begin{figure}
\includegraphics[width=18cm]{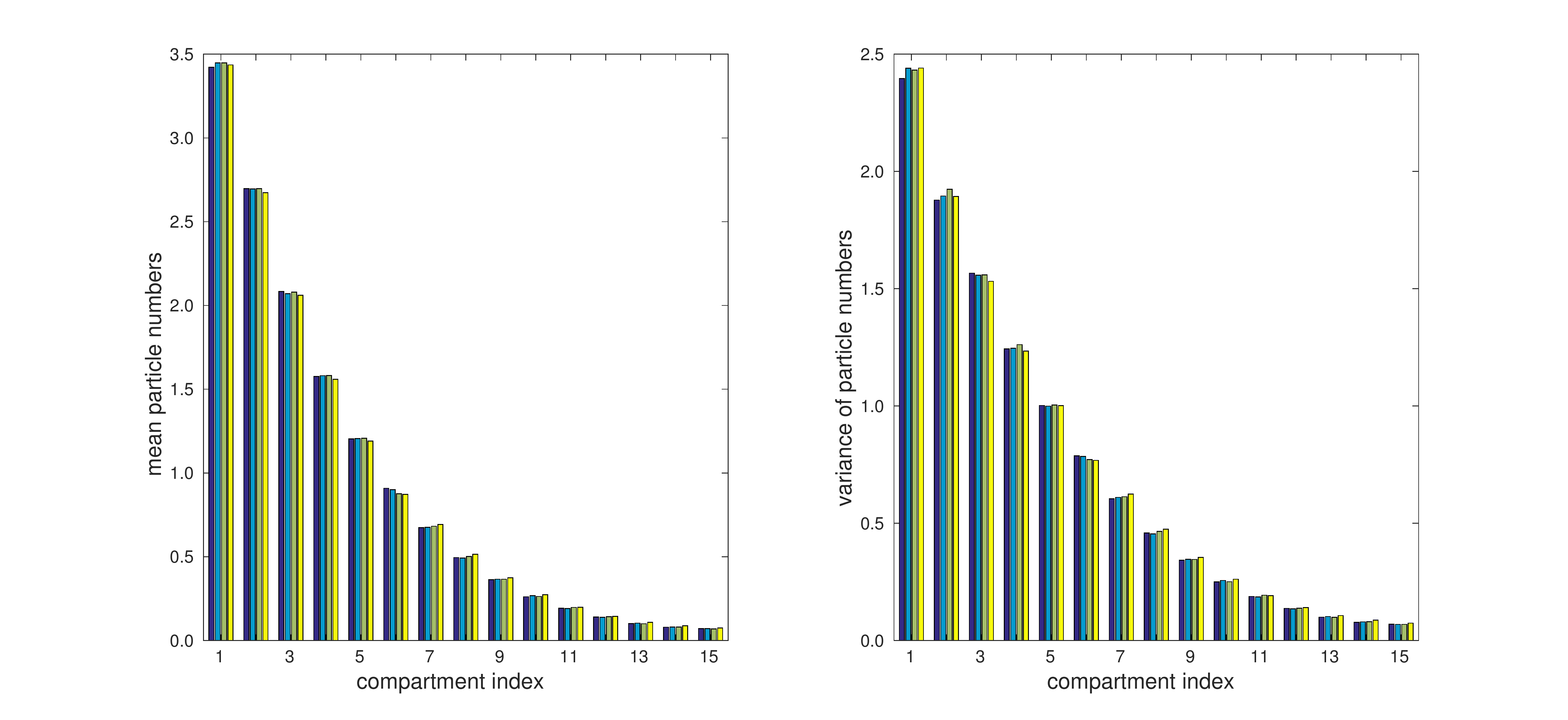}
\caption{Results from 50,000 repeats of the morphogen gradient simulation at time $t=25$. Particles are added to the system with rate $r_1=1$, and are placed into the closest compartment to the left-hand boundary which is not full to capacity, while each particle may decay and be removed with rate $r_2=0.05$. From left to right, each set of four bars shows results from the fully-excluding (dark blue), partially-excluding (light blue), Voronoi (green) and pseudo-compartment (yellow) models.}
\label{fig:MorphogenGrad_MeanAndVar}
\end{figure}

The means and variances of particles at time $t=25$ are illustrated in Figure \ref{fig:MorphogenGrad_MeanAndVar}, while HDE values and the computational time required to simulate each model are presented in Table \ref{table:morphogengradient}. In this case, the hybrid simulations are significantly faster than the equivalent fully-excluding simulations. This is because the majority of particles are located in the first third of the domain, where the hybrid models use computationally efficient partially-excluding compartments. Simulations of the Voronoi model ran more than four times faster than the fully-excluding model, while simulations of the pseudo-compartment model ran more than three times faster than the fully-excluding model.

\begin{center}
\begin{table}
\caption{Details of morphogen gradient simulation.}
    \begin{tabular}{ | l || P{20mm} | P{25mm} | P{35mm} | P{40mm} |}
    \hline
    Model 			& Mean HDE& Variance HDE & Time to simulate & Acceleration relative \\
				& at $t=25$& at $t=25$ &  &  to fully-excluding	\\ \hline \hline
    Fully-excluding 	& 0.0015 & 0.0050 & 16622 seconds & - \\ \hline%4 hours, 37 minutes, 2 seconds  \\ \hline
    Partially-excluding & 0.0023 & 0.0060 & 385 seconds & 43.2 times faster \\ \hline%6 minutes, 25 seconds  \\ \hline
    Voronoi 		& 0.0032 & 0.0059 & 3438 seconds & 4.8 times faster \\ \hline%57 minutes, 18 seconds  \\ \hline
    Pseudo-compartment  & 0.0077 & 0.0105 & 4847 seconds & 3.4 times faster \\ \hline %1 hours, 20 minutes, 47 seconds  \\ \hline
    \end{tabular}
    \label{table:morphogengradient}
\end{table}
\end{center}

%\begin{eqnarray}
% \mathrm{Coarse Mean}&=&0.0061,\nonumber\\
% \mathrm{Voronoi Mean}&=&0.0067,\nonumber\\
% \mathrm{Pseudo-compartment Mean}&=&0.0074,\nonumber\\
% &&\nonumber\\
% \mathrm{Coarse Variance}&=&0.0058,\nonumber\\
% \mathrm{Voronoi Variance}&=&0.0052,\nonumber\\
% \mathrm{Pseudo-compartment Variance}&=&0.0071.\nonumber
%\end{eqnarray}

%Time to simulate: fully-excluding took 3 hours, 2 minutes, 2 seconds; partially-excluding took 6 minutes, 11 seconds; Voronoi took 14 minutes, 33 seconds; pseudo-compartment took 22 minutes, 14 seconds.

%\begin{figure}
%\includegraphics[width=16cm]{MorphogenGrad_Mean_2015_11_11.jpg}
%\caption{Results from 50,000 repeats of morphogen gradient simulation. Results taken from final state at time $t=25$. Particles are added to the system with propensity 1, and are placed into the closest compartment to the left-hand boundary which is not full to capacity, while each particle may decay and be removed with propensity $0.05$.}
%\label{fig:MorphogenGradient_Mean}
%\end{figure}

%\begin{figure}
%\includegraphics[width=16cm]{MorphogenGrad_Var_2015_11_11.jpg}
%\caption{Same data as Fig \ref{fig:MorphogenGradient_Mean}, but showing variance values.}
%\label{fig:MorphogenGradient_Var}
%\end{figure}
 
 \section{Application to a simple multi-species exclusion system}
\label{sec:SimpleMultiSpeciesExclusionSystem}
Having established the accuracy of the hybrid methodologies in the test cases of the previous section, we now apply them to a simple model system where the results of the partially-excluding model deviate from those of the fully-excluding model.

\begin{figure}
\includegraphics[width=14cm]{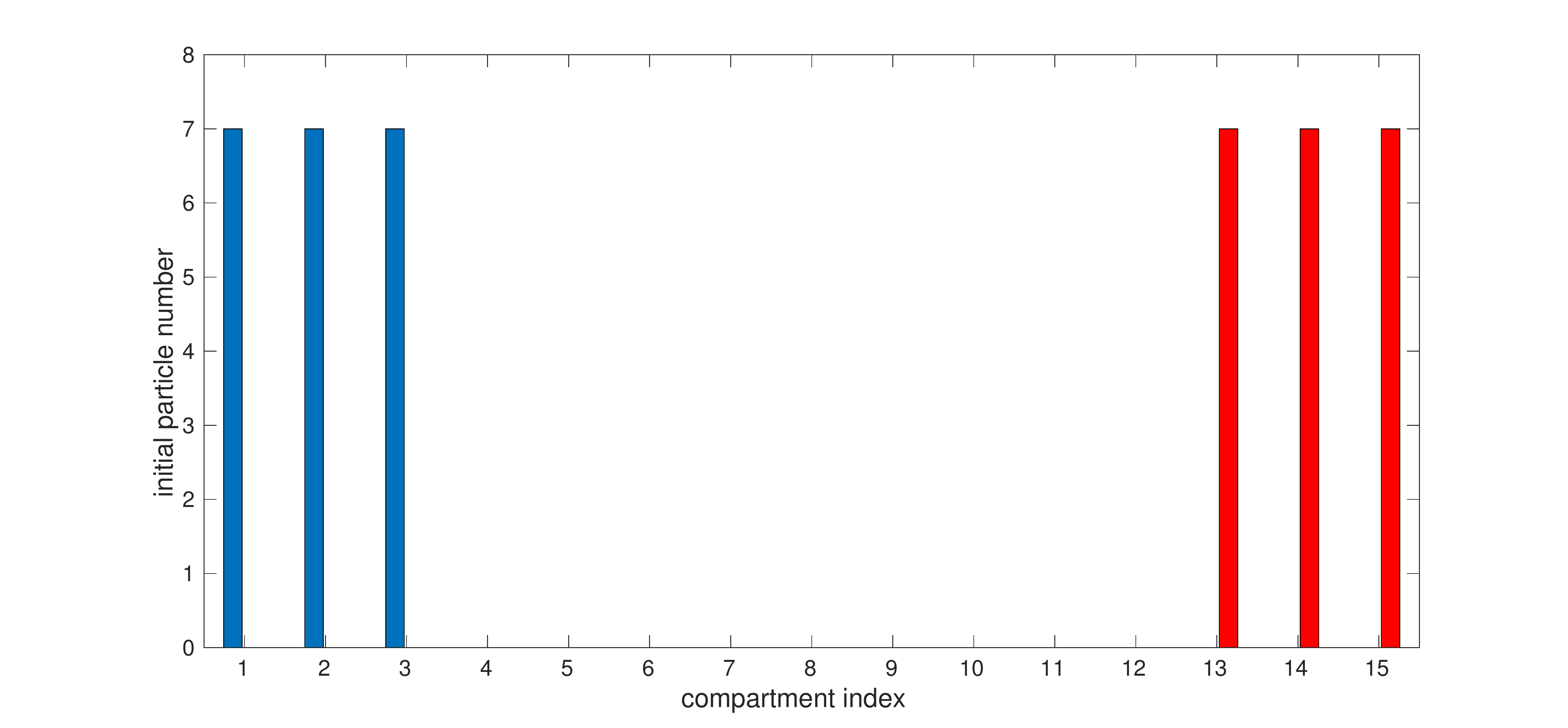}
\caption{At time $t=0$, twenty-one particles of species A (blue) are initialised at the left of the domain, and twenty-one particles of species B (red) are initialised at the right.}
\label{fig:MultispeciesInitialState}
\end{figure}

We consider a simple multi-species system containing two species of particle, A and B~\cite{Simpson:2009:MSE}. Neither species is reactive, and particles do not interact with each other except through volume exclusion effects. The initial state for the system consists of twenty-one particles of species A at the left of the domain, and twenty-one particles of species B at the right of the domain, as illustrated in Figure \ref{fig:MultispeciesInitialState} (so that for the fully-excluding model, the first and last twenty-one compartments are occupied, and for the other models the first and last three compartments are filled to capacity).
When simulated using a partially-excluding model, this system will eventually reach a homogeneous steady state, with both A and B particles intermixed since the coarsened lattice description allows particles to move past each another. In contrast, the fully-excluding model does not allow particles to pass one another, and hence the two species will remain separated. Our hybrid methods can improve the efficiency of simulations by modelling a small central region, where it is possible for the two species to meet, using a fully-excluding lattice, while using a computationally efficient, partially-excluding lattice elsewhere in the domain.

\begin{figure}
\includegraphics[width=12cm]{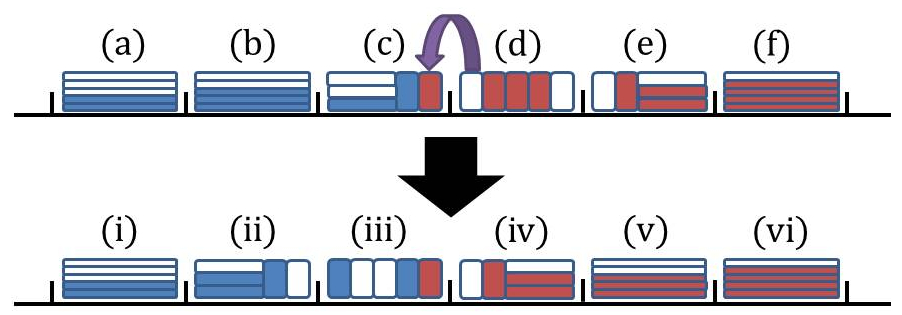}
\caption{An illustration of an adaptive Voronoi domain partition. Particles of species A and B are shown in blue and red, respectively. The upper image shows the lattice before the hybrid region shifts left, where the areas marked (a), (b) and (f) are partially-excluding, those marked (c) and (e) are a combination of fully-excluding and partially-excluding, and the one marked (d) is a section of fully-excluding compartments. When a B particle enters box (c), the interface shifts left as illustrated in the lower image, where the areas marked (i), (v) and (vi) are partially-excluding, those marked (ii) and (iv) are a combination of fully-excluding and partially-excluding, and the one marked (iii) is a section of fully-excluding compartments. Where a partially-excluding box is replaced with a finer partition (i.e. (b) and the left part of (c)), the particles within it are uniformly randomised across the compartments taking its place, weighted by the length of the destination compartments. When a section of compartments are replaced by a coarser partition (i.e. (e) and the right of (d)) the occupancy of the new compartments is found from the total number of particles across that section.} 
\label{fig:AdaptiveModel_ShiftingVoronoiInterface_Cropped}
\end{figure}

\begin{figure}
\includegraphics[width=12cm]{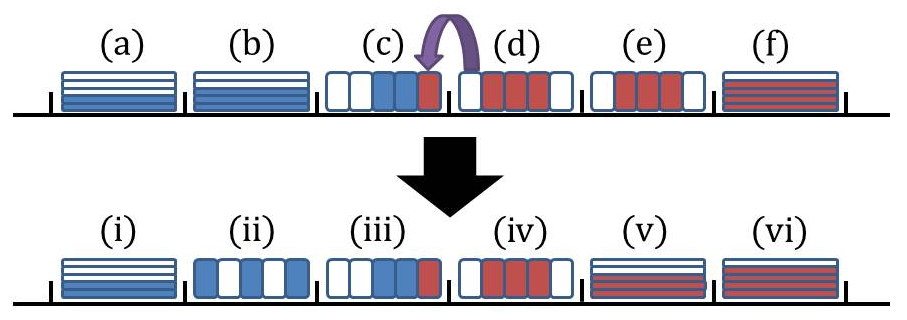}
\caption{An illustration of an adaptive domain partition using the pseudo-compartment model. Particles of species A and B are shown in blue and red, respectively. The upper image shows the lattice before the hybrid region shifts left, where the areas marked (a), (b) and (f) are partially-excluding, those marked (c) and (e) are pseudo-compartments, and the one marked (d) is a section of fully-excluding compartments. When a B particle enters box (c), the interface shifts left as illustrated in the lower image, where the areas marked (i), (v) and (vi) are partially-excluding, those marked (ii) and (iv) are pseudo-compartments, and the one marked (iii) is a section of fully-excluding compartments. The positions of any particles in box (b) are uniformly randomised across (ii), while the occupancy of (v) is given by the total number of particles across (e).}
\label{fig:AdaptiveModel_ShiftingPseudoInterface_Cropped}
\end{figure}

Figures \ref{fig:AdaptiveModel_ShiftingVoronoiInterface_Cropped} and \ref{fig:AdaptiveModel_ShiftingPseudoInterface_Cropped} illustrate how such regions can be arranged for Voronoi and pseudo-compartment models where $m=5$, and how the interfaces can be shifted left or right to prevent particles of species A and B from passing one another. Shifting the interface imposes a small computational cost, but this is more than compensated for by the decreased number of jump events occurring due to the coarser lattice partitioning. \red{This approach could be generalised to a model containing additional particle species by considering additional fully-excluding regions between each species (assuming all particles of each species are initialised in distinct groups). In cases where multiple species are already mixed together at time $t=0$, but preserving their ordering is important for the model, it may be necessary to employ a fully-excluding model.} Although it would not be necessary to prevent particle mixing in this way in a two- or three-dimensional model, it would still be possible to dynamically adjust the lattice in similar ways to save computational time where possible and preserve detailed modelling where necessary~\cite{Bayati:2011:AMR}.

We adopt the same values of $N$, $h$, $L$ and $D$ used in Section \ref{section:NumericalInvestigations}, and perform $50,000$ realisations of each model over the time interval $t\in[0,25]$. For the Voronoi and pseudo-compartment models, we begin each realisation with six compartments of capacity $m=7$ at the left-hand side and at the right-hand side of the domain, with an intermediate region at the centre of the domain of the form illustrated in Figures \ref{fig:AdaptiveModel_ShiftingVoronoiInterface_Cropped} and \ref{fig:AdaptiveModel_ShiftingPseudoInterface_Cropped}.

\begin{figure}
\includegraphics[width=18cm]{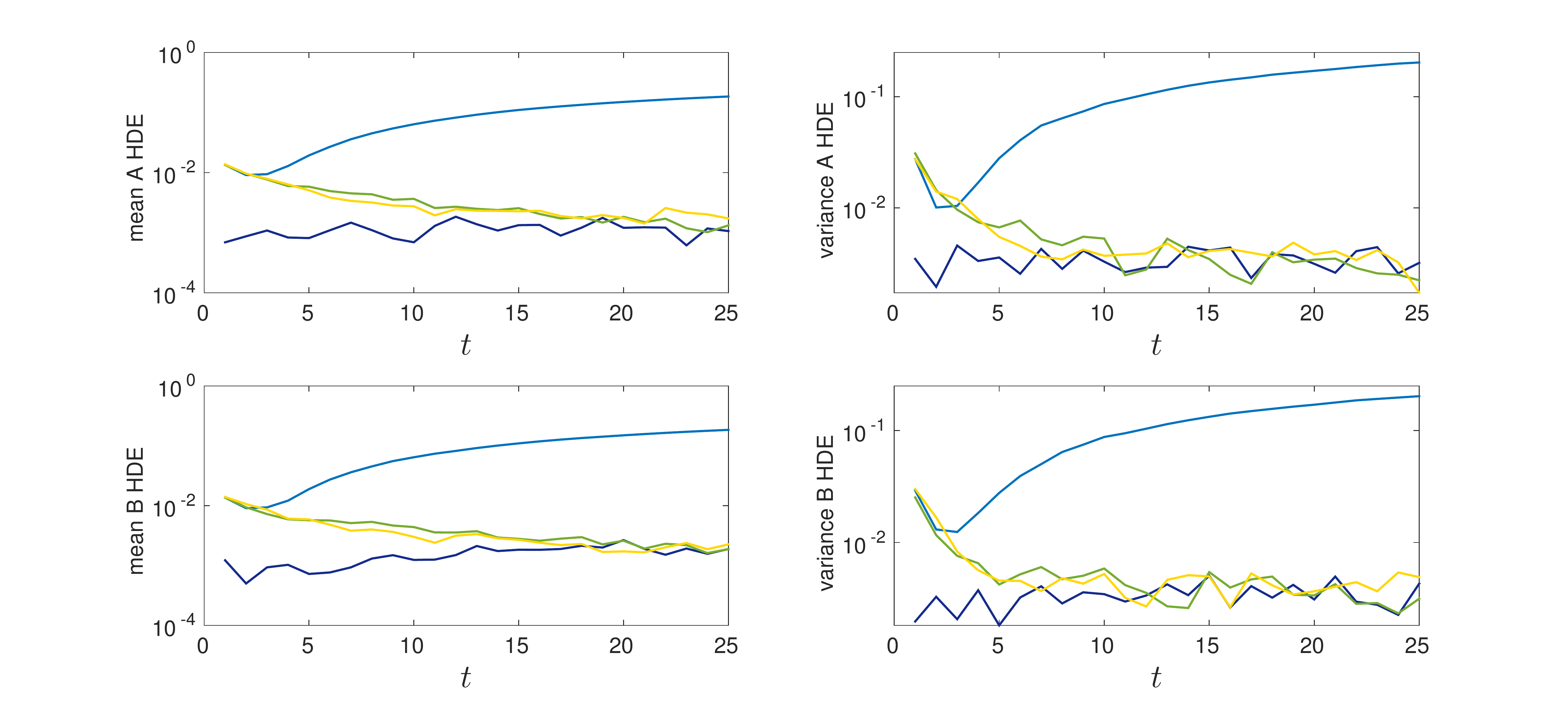}
\caption{HDE values for the mean and variance of the multi-species simulation results at discrete times $t=1,2,\dots,25$. The growing deviation of the partially-excluding model (plotted in light blue) from the other three models (fully-excluding, Voronoi, and pseudo-compartment, plotted in dark blue, green and yellow, respectively) can be clearly seen in each case.}
\label{fig:AdaptiveMultispecies_HDE_50k}
\end{figure}

\begin{figure}
\includegraphics[width=18cm]{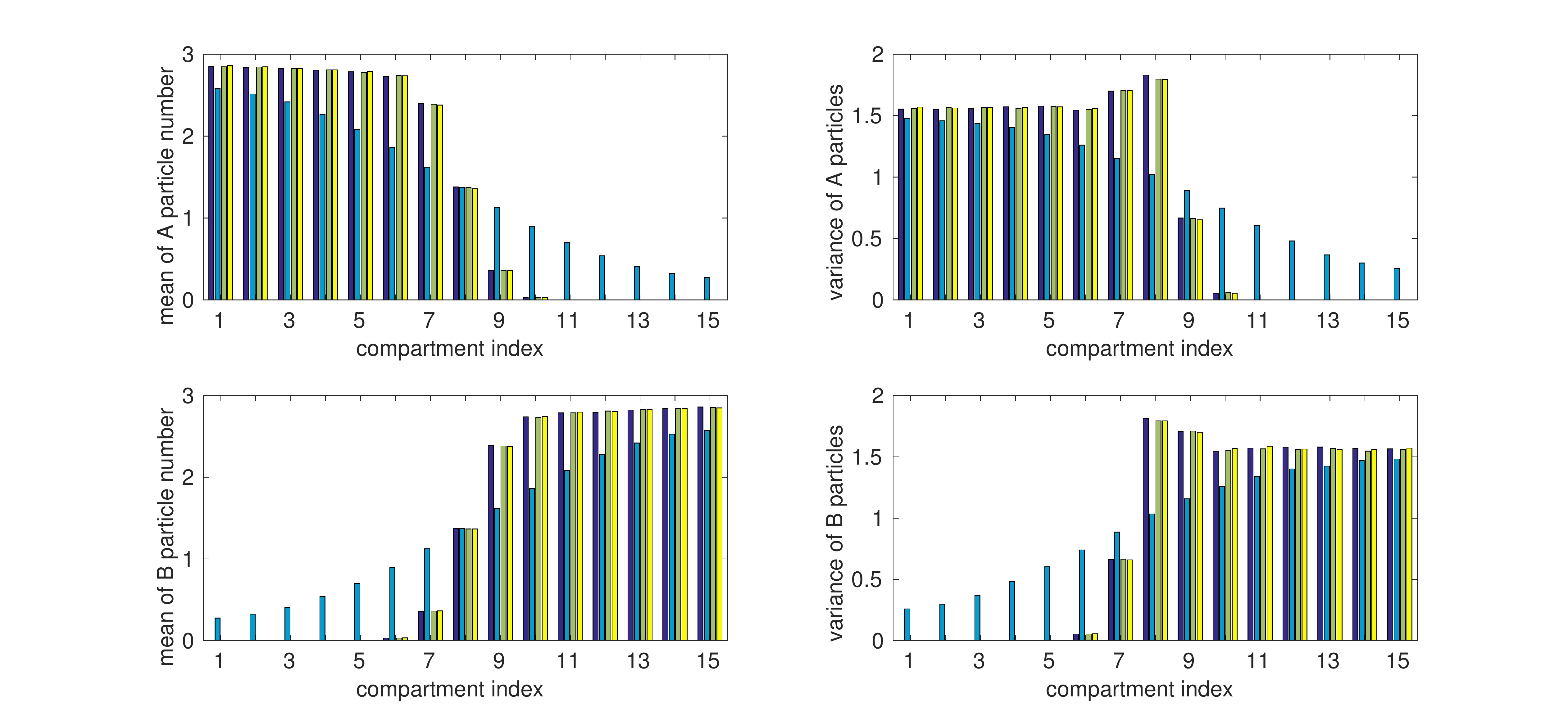}
\caption{Results from $50,000$ repeats of the multi-species simulation at time $t=25$. From left to right, each set of four bars shows results from the fully-excluding (dark blue), partially-excluding (light blue), Voronoi (green) and pseudo-compartment (yellow) models. The fully-excluding and hybrid model results display strong agreement, but the results of the partially-excluding model differ as this model cannot prevent the particles of differing type from moving past one another.}
\label{fig:AdaptiveMultispecies_ATop_BBottom_50k}
\end{figure}

We compare mean and variance values at $t=1,2,3\dots,25$, and plot the resulting HDEs in Figure \ref{fig:AdaptiveMultispecies_HDE_50k}. As anticipated, the results generated using the partially-excluding model deviate significantly from those generated using the fully-excluding and hybrid models. This deviation is illustrated by plots of the means and variances of particle numbers at $t=25$ in Figure \ref{fig:AdaptiveMultispecies_ATop_BBottom_50k}. The HDE values at time $t=25$ are listed in Table \ref{table:MultiSpecies} together with the time taken to simulate each system. It can be seen that the Voronoi and pseudo-compartment models run 6.9 and 3.2 times faster, respectively, than the fully-excluding system without losing accuracy. 

 \begin{center}
\begin{table}
\caption{Details of multi-species simulation.}
    \begin{tabular}{ | l || P{27.5mm} | P{27.5mm} | P{30mm} | P{35mm} |}
    \hline
    Model 		& Mean HDE  		& Variance HDE 	  	& Time to simulate 	& Acceleration relative \\
			& at $t=25$ (A/B)	& at $t=25$ (A/B) 	&			& to fully-excluding	\\ \hline \hline
    Fully-excluding 	& 0.0011/0.0019 	& 0.0032/0.0043 	& 53663 seconds 		& - 			\\ \hline %7 hours, 59 minutes, 1 seconds  \\ \hline
    Partially-excluding & 0.1853/0.1844 	& 0.2034/0.2023 	& 902 seconds		& 59.5 times faster	\\ \hline %9 minutes, 42 seconds  \\ \hline
    Voronoi 		& 0.0013/0.0019 	& 0.0022/0.0031 	& 7722 seconds 		& 6.9 times faster 	\\ \hline %5 hours, 28 minutes, 58 seconds  \\ \hline
    Pseudo-compartment 	& 0.0017/0.0023 	& 0.0017/0.0049 	& 16755 seconds 		& 3.2 times faster 	\\ \hline %7 hours, 19 minutes, 21 seconds  \\ \hline
    \end{tabular}
    \label{table:MultiSpecies}
\end{table}
\end{center}
  
\section{Discussion}

In this paper, we have presented two methods for the simulation of volume-excluding compartment-based models on non-uniform lattices. We have presented analytical arguments for the agreement between aggregated mean particle distribution in the uniform partially-excluding model with $m>1$ and in both hybrid models (i.e. $M^{(m)}$ $\mu^{(v)}$ and $\mu^{(p)}$, respectively) with the aggregated mean particle distribution in the `accurate' fully-excluding model with $m=1$ ($\mu^{(m)}$). Numerical investigations of three single-species test cases confirm that all four models agree on the mean and variance of particle numbers within each compartment. We have further presented a simple multi-species system, where the results of uniform partially-excluding models deviate from those of the accurate fully-excluding model, but the results of hybrid models match those of the fully-excluding model and can be generated in a fraction of the time. This is a valuable development for researchers working with multi-scale diffusion systems, as it enables computationally efficient partially-excluding compartments to be used where possible, while using fine-grained fully-excluding compartments elsewhere in the domain when necessary. Time savings of nearly a factor of seven relative to the fully-excluding model were observed for the Voronoi model when the test system was conducive to hybrid modelling, i.e. when the majority of particles spend most of the simulation within the coarse-grained region. The time savings in other situations could be even more significant. Partially-excluding models are generally accurate, and are consistently faster than both fully-excluding and hybrid models, but the multi-species example in Section VI shows that they can be inadequate when fine spatial resolution is required in a region of the domain. In future work, we will consider two-dimensional hybrid models in greater detail, and discuss the implementation of reactions between particles in this framework.

\begin{acknowledgments}
 PRT gratefully acknowledges an EPSRC studentship through The University of Oxford's Systems Biology Doctoral Training Centre. MJS acknowledges support from the Australian Research Council (FT130100148). The authors thank two anonymous reviewers for helpful comments on an earlier version of this manuscript.
\end{acknowledgments}

\newpage

\appendix

\section{Outline of transition rate derivations}
\label{app:VoronoiJumpDerivations}

Transition rates between compartments can be obtained from standard first passage process results~\cite{Yates:2012:GMM}. Consider a particle undergoing Brownian motion, starting from $x_j$, the residence point of the $j^{th}$ compartment. The probability distribution for its position, $p(x,t)$, evolves in time following the diffusion equation,
\begin{equation}
 \dfrac{\partial p}{\partial t}=D\dfrac{\partial^2 p}{\partial x^2}.
\end{equation}
To calculate transition rates between compartments, we consider a first passage process on the interval $x\in[x_{j-1},x_{j-1}]$ to determine the expected time until a particle starting from one residence point reaches one of its neighbouring residence points. For the diffusion equation, this means defining boundary conditions $p(x_{j-1},t)=p(x_{j-1},t)=0$, and initial condition $p(x,0)=\delta(x-x_{j})$. By applying a Laplace transform, it is possible to obtain the eventual hitting probabilities describing the likelihood of the particle leaving to either side of the interval~\cite{Redner:2001:FPP}
\begin{eqnarray}
 \epsilon_{-}(x_{j})&=&\dfrac{x_{j+1}-x_{j}}{x_{j+1}-x_{j-1}},\\
 \epsilon_{+}(x_{j})&=&\dfrac{x_{j}-x_{j-1}}{x_{j+1}-x_{j-1}}.
\end{eqnarray}
Combining these values with the conditional mean exit times
\begin{eqnarray}
 \langle t(x_{j})\rangle_{-}&=&\dfrac{(x_{j}-x_{j-1})(2x_{j+1}-x_{j}-x_{j-1})}{6D},\\
 \langle t(x_{j})\rangle_{+}&=&\dfrac{(x_{j+1}-x_{j})(x_{j+1}+x_{j}-2x_{j-1})}{6D},
\end{eqnarray}
we can then obtain the unconditional exit time
\begin{eqnarray}
 \langle t(x_{j})\rangle&=&\epsilon_{-}(x_{j})\langle t(x_{j})\rangle_{-}+\epsilon_{+}(x_{j})\langle t(x_{j})\rangle_{+}\nonumber\\
 &=&\dfrac{(x_{j}-x_{j-1})(x_{j+1}-x_{j})\left([2x_{j+1}-x_{j}-x_{j-1}]+[x_{j+1}+x_{j}-2x_{j-1}]\right)}{6D(x_{j+1}-x_{j-1})}\nonumber\\
 &=&\dfrac{(x_{j}-x_{j-1})(x_{j+1}-x_{j})\left(3x_{j+1}-3x_{j-1}\right)}{6D(x_{j+1}-x_{j-1})}\nonumber\\
 &=&\dfrac{(x_{j}-x_{j-1})(x_{j+1}-x_{j})}{2D}.
\end{eqnarray}
Inverting the unconditional exit times gives the unconditional exit rate, and by multiplying this value by the hitting probabilities we obtain the \red{non-excluding} transition rates
\begin{eqnarray}
 \red{\mathcal{T}}_{j}^{-}&=&\dfrac{\epsilon_{-}(x_{j})}{\langle t(x_{j})\rangle}=\dfrac{2D}{(x_{j}-x_{j-1})(x_{j+1}-x_{j-1})}=\dfrac{Dh}{\Delta x_{j}m_{j}},\nonumber\\
 \red{\mathcal{T}}_{j}^{+}&=&\dfrac{Dh}{\Delta x_{j+1}m_{j}}.
\end{eqnarray}
It is also possible to derive these values using a finite element approach~\cite{Engblom:2009:SUM}. Transition rates for the two boundary compartments, $j=1,K$, can be derived similarly using the reflection principle of Brownian motion. For example, define a notional point $x_{0}=-x_{2}$, then the unconditional expected exit time for the first box is
\begin{eqnarray}
 \langle t(x_{1})\rangle&=&\dfrac{(x_{1}-x_{0})(x_{2}-x_{1})}{2D}=\dfrac{(x_{1}+x_{2})(x_{2}-x_{1})}{2D}.
\end{eqnarray}
Clearly $\epsilon_{+}(x_{1})=1$, hence we write
\begin{equation}
 \red{\mathcal{T}}_{1}^{+}=\dfrac{Dh}{\Delta x_{2}m_{1}}.
\end{equation}
We can similarly derive $\red{\mathcal{T}}_{K}^{-}$ by reflecting $x_{K-1}$ about $x=L$.

\section{Voronoi master equation derivations}

\label{appendix:VoronoiMasterEquationDerivation}

The following derivations are based upon those found in the Supplemental Information of a previous paper, generalised here to a non-uniform Voronoi lattice~\cite{Taylor:2015:RTM}. Recall that a spatial lattice composed of $K$ compartments has been defined. The distribution of particles over this domain is given by the vector $\textbf{n}(t)=[n_1(t),n_2(t),...,n_K(t)]$, where $n_i(t)$ denotes the number of particles in the $i^{\mathrm{th}}$ compartment at time $t$. Particles in compartment $i$ may \textit{attempt} to jump out to compartments $i-1$ or $i+1$ with rates $\red{\mathcal{T}}_i^-$ and $\red{\mathcal{T}}_i^+$, respectively.

We define two operators, $J_{i}^{+}:\mathbb{R}^{K}\rightarrow\mathbb{R}^{K}$, for $i=1,...,K-1$, and 
$J_{i}^{-}:\mathbb{R}^{K}\rightarrow\mathbb{R}^{K}$, for $i=2,...,K$, as
\begin{eqnarray}
&J_{i}^{+}:[n_1,...,n_i,...,n_K] \rightarrow [n_1,...,n_{i-2},n_{i-1},n_{i}+1,n_{i+1}-1,n_{i+2},...,n_K], \\
&J_{i}^{-}:[n_1,...,n_i,...,n_K] \rightarrow [n_1,...,n_{i-2},n_{i-1}-1,n_{i}+1,n_{i+1},n_{i+2},...,n_K].
\end{eqnarray}
Both operators move a particle into compartment $i$, taken from the compartment to the right or left, respectively. We assume that attempted jumps into some compartment $j$ fail with probability $n_j/m_j$ due to exclusion effects. We can then write the probability master equation as
\begin{eqnarray}
\frac{\textrm{d} \mathrm{Pr}(\textbf{n},t)}{\textrm{d} t}=&\sum\limits_{i=1}^{K-1}&\red{T}_i^+\left\{(n_i+1)\mathrm{Pr}(J_i^+\textbf{n},t)-n_i\mathrm{Pr}(\textbf{n},t)\right\}\nonumber\\
+&\sum\limits_{i=2}^{K}&\red{T}_i^-\left\{(n_i+1)\mathrm{Pr}(J_i^-\textbf{n},t)-n_{i}\mathrm{Pr}(\textbf{n},t)\right\}.\nonumber\\
=&\sum\limits_{i=1}^{K-1}&\red{\mathcal{T}}_i^+\left\{(n_i+1)\left[1-\dfrac{n_{i+1}-1}{m_{i+1}}\right]\mathrm{Pr}(J_i^+\textbf{n},t)-n_i\left[1-\dfrac{n_{i+1}}{m_{i+1}}\right]\mathrm{Pr}(\textbf{n},t)\right\} \nonumber\\
\label{eq:lastprobequationbeforemeans}+&\sum\limits_{i=2}^{K}&\red{\mathcal{T}}_i^-\left\{(n_i+1)\left[1-\dfrac{n_{i-1}-1}{m_{i-1}}\right]\mathrm{Pr}(J_i^-\textbf{n},t)-n_i\left[1-\dfrac{n_{i-1}}{m_{i+1}}\right]\mathrm{Pr}(\textbf{n},t)\right\}.\nonumber\\
\end{eqnarray}
Define the mean vector, $\textbf{M}=[M_1,...,M_K]$, where
\begin{equation}
M_j=\sum\limits_{n_1=1}^{N}\sum\limits_{n_2=1}^{N}\dots\sum\limits_{n_K=1}^{N}n_j\mathrm{Pr}(\textbf{n},t):=\sum\limits_{n_1,n_2,...,n_K=0}^{\mathcal{N}}n_j\mathrm{Pr}(\textbf{n},t).
\end{equation}
Multiplying Eq. \eqref{eq:lastprobequationbeforemeans} by $n_j$ and summing over all possible values that the vector $\textbf{n}(t)$ can take we have
\begin{eqnarray}
\label{eq:PME_start}
\frac{\textrm{d} M_j}{\textrm{d} t}&=&\sum\limits_{n_1,n_2,...,n_K=0}^{\mathcal{N}}n_j\left(\sum\limits_{i=1}^{K-1}\red{\mathcal{T}}_i^+\left\{(n_i+1)\left[1-\dfrac{n_{i+1}-1}{m_{i+1}}\right]\mathrm{Pr}(J_i^+\textbf{n},t)-n_i\left[1-\dfrac{n_{i+1}}{m_{i+1}})\right]\mathrm{Pr}(\textbf{n},t)\right\}\right. \nonumber\\
&&\qquad\qquad\left.+\sum\limits_{i=2}^{K}\red{\mathcal{T}}_i^-\left\{(n_i+1)\left[1-\dfrac{n_{i-1}-1}{m_{i+1}}\right]\mathrm{Pr}(J_i^-\textbf{n},t)-n_i\left[1-\dfrac{n_{i-1}}{m_{i+1}}\right]\mathrm{Pr}(\textbf{n},t)\right\}\right).\nonumber\\
\end{eqnarray}
We begin by considering only the first term of this expression, with $\mathrm{Pr}(J_i^{+}\textbf{n},t)$ expanded explicitly to give
\begin{equation}
 \sum\limits_{n_1,n_2,...,n_K=0}^{\mathcal{N}}n_j\sum\limits_{i=1}^{K-1}d(n_i+1)\left[1-f(n_{i+1}-1)\right]\mathrm{Pr}(n_1,n_2,...,n_i+1,n_{i+1}-1,...n_K,t).
\end{equation}
When $i\neq j,j-1$ each term of this expression reduces to
\begin{equation}
\red{\mathcal{T}}_i^+\langle n_jn_i\rangle-\red{\mathcal{T}}_i^+\langle n_jn_if(n_{i+1})\rangle,
\end{equation}
where $\langle n_jn_i\rangle$ indicates the mean of the product. When $i=j$,
\begin{eqnarray}
&\sum\limits_{n_1,n_2,...,n_K=0}^{\mathcal{N}}n_{j}\red{\mathcal{T}}_{j}^{+}(n_j+1)\left[1-f(n_{j+1}-1)\right]\mathrm{Pr}(n_1,n_2,...,n_j+1,n_{j+1}-1,...n_K,t) \nonumber\\
=&\sum\limits_{n_1,n_2,...,n_K=0}^{\mathcal{N}}\red{\mathcal{T}}_{j}^{+}(n_j+1)^2\left[1-f(n_{j+1}-1)\right]\mathrm{Pr}(n_1,n_2,...,n_j+1,n_{j+1}-1,...n_K,t) \nonumber\\
&-\red{\mathcal{T}}_{j}^{+}(n_j+1)\left[1-f(n_{j+1}-1)\right]\mathrm{Pr}(n_1,n_2,...,n_j+1,n_{j+1}-1,...n_K,t)\nonumber\\
=&\red{\mathcal{T}}_{j}^{+}[\langle n_jn_j\rangle-\langle n_jn_jf(n_{j+1})\rangle-M_j+\langle n_jf(n_{j+1})\rangle].
\end{eqnarray}
Similarly for $i=j-1$,
\begin{eqnarray}
&\sum\limits_{n_1,n_2,...,n_K=0}^{\mathcal{N}}\red{\mathcal{T}}_{j-1}^{+}n_j(n_{j-1}+1)\left[1-f(n_{j}-1)\right]\mathrm{Pr}(n_1,n_2,...,n_{j-1}+1,n_{j}-1,...n_K,t)\nonumber\\
=&\red{\mathcal{T}}_{j-1}^{+}[\langle n_{j-1}n_j\rangle-\langle n_{j-1}n_jf(n_j)\rangle+M_{j-1}-\langle n_{j-1}f(n_j)\rangle].
\end{eqnarray}
We then consider the second term,
\begin{equation}
-\sum\limits_{n_1,n_2,...,n_K=0}^{\mathcal{N}}n_j\sum\limits_{i=1}^{K-1}\red{\mathcal{T}}_{i}^{+}n_i\left[1-f(n_{i+1})\right]\mathrm{Pr}(n_1,n_2,...,n_i,...,n_K,t),
\end{equation}
which evaluates to
\begin{equation}
-\sum\limits_{i=1}^{K-1} \red{\mathcal{T}}_{i}^{+}\left[\langle n_in_j\rangle-\langle n_if(n_{i+1})n_j\rangle\right].
\end{equation}
When combined with the expressions derived from the first term these give us
\begin{equation}\label{eq:firsthalfofmeanPME}
 -\red{\mathcal{T}}_{j}^{+}M_j+\red{\mathcal{T}}_{j}^{+}\langle n_{j}f(n_{j+1})\rangle+\red{\mathcal{T}}_{j-1}^{+}M_{j-1}-\red{\mathcal{T}}_{j-1}^{+}\langle n_{j-1}f(n_j) \rangle.
\end{equation}
Applying the same approach to the third and fourth terms we obtain
\begin{equation}
 -\red{\mathcal{T}}_{j}^{-}M_j+\red{\mathcal{T}}_{j}^{-}\langle n_{j}f(n_{j-1})\rangle+\red{\mathcal{T}}_{j+1}^{-}M_{j+1}-\red{\mathcal{T}}_{j+1}^{-}\langle n_{j+1}f(n_j) \rangle.
\end{equation}
Adding this expression to Eq. (\ref{eq:firsthalfofmeanPME}), we arrive at
\begin{eqnarray}
 \dfrac{\mathrm{d}M_j}{\mathrm{d}t}=&&\red{\mathcal{T}}^{+}_{j-1}\left(M_{j-1}-\dfrac{1}{m_j}\langle n_{j-1}n_j\rangle\right)-\red{\mathcal{T}}^{-}_{j}\left(M_{j}-\dfrac{1}{m_{j-1}}\langle n_{j-1}n_j\rangle\right)\nonumber\\
 &-&\red{\mathcal{T}}^{+}_{j}\left(M_{j}-\dfrac{1}{m_{j+1}}\langle n_{j}n_{j+1}\rangle\right)+\red{\mathcal{T}}^{-}_{j+1}\left(M_{j+1}-\dfrac{1}{m_{j}}\langle n_{j}n_{j+1}\rangle\right),
 \end{eqnarray}
as stated in Eq. (\ref{eq:InitialMeanEvolutionEquationWithNonlinTerms}) of the main text.

Variance equations can be derived using the same approach. We begin by using Eq. (\ref{eq:lastprobequationbeforemeans}) to obtain
\begin{eqnarray}
 \dfrac{\mathrm{d}}{\mathrm{d}t}\sum\limits^{\mathcal{N}}n_i^2P(\textbf{n})=&\red{\mathcal{T}}^+_{i-1}&\left(2\langle n_{i-1}n_{i}\rangle-\dfrac{2}{m_i}\langle n_{i-1}n_{i}n_{i}\rangle+M_{i-1}-\dfrac{1}{m_i}\langle n_{i-1}n_{i}\rangle \right)\nonumber\\
 &+\red{\mathcal{T}}^+_{i}&\left(-2\langle n_{i}n_{i}\rangle+\dfrac{2}{m_{i+1}}\langle n_{i}n_{i}n_{i+1}\rangle+M_{i}-\dfrac{1}{m_{i+1}}\langle n_{i}n_{i+1}\rangle \right)\nonumber\\
 &+\red{\mathcal{T}}^-_{i}&\left(-2\langle n_{i}n_{i}\rangle+\dfrac{2}{m_{i-1}}\langle n_{i-1}n_{i}n_{i}\rangle+M_{i}-\dfrac{1}{m_{i-1}}\langle n_{i-1}n_{i}\rangle \right)\nonumber\\
 &+\red{\mathcal{T}}^-_{i+1}&\left(2\langle n_{i}n_{i+1}\rangle-\dfrac{2}{m_{i}}\langle n_{i}n_{i}n_{i+1}\rangle+M_{i+1}-\dfrac{1}{m_{i}}\langle n_{i}n_{i+1}\rangle \right)\nonumber,\\
 \Rightarrow\dfrac{\mathrm{d}}{\mathrm{d}t}\langle n_{i}^2\rangle=&\red{\mathcal{T}}^+_{i-1}&\left(2\langle n_{i-1}n_{i}\rangle+M_{i-1}-\dfrac{1}{m_i}\langle n_{i-1}n_{i}\rangle \right)\nonumber\\
 &+\red{\mathcal{T}}^+_{i}&\left(-2\langle n_{i}n_{i}\rangle+M_{i}-\dfrac{1}{m_{i+1}}\langle n_{i}n_{i+1}\rangle \right)\nonumber\\
 &+\red{\mathcal{T}}^-_{i}&\left(-2\langle n_{i}n_{i}\rangle+M_{i}-\dfrac{1}{m_{i-1}}\langle n_{i-1}n_{i}\rangle \right)\nonumber\\
 &+\red{\mathcal{T}}^-_{i+1}&\left(2\langle n_{i}n_{i+1}\rangle+M_{i+1}-\dfrac{1}{m_{i}}\langle n_{i}n_{i+1}\rangle \right),
\end{eqnarray}
where we have used $\red{\mathcal{T}}^{-}_{i}/m_{i-1}=\red{\mathcal{T}}^{+}_{i-1}/m_{i}$ and $\red{\mathcal{T}}^{-}_{i+1}/m_{i}=\red{\mathcal{T}}^{+}_{i}/m_{i+1}$. We then use $\langle n_{i}n_{j}\rangle=V_{i,j}-M_{i}M_{j}$ to obtain
\begin{eqnarray}
 \dfrac{\mathrm{d}V_j}{\mathrm{d}t}&=&-2\left(\red{\mathcal{T}}^-_{j}+\red{\mathcal{T}}^+_{j}\right)V_j+2T^+_{j-1}\left(1-\dfrac{1}{m_j}\right)V_{j-1,j}+2T^-_{j+1}\left(1-\dfrac{1}{m_j}\right)V_{j,j+1}\nonumber\\
 &&+\red{\mathcal{T}}^+_{j-1}M_{j-1}\left(1-\dfrac{M_{j}}{m_{j}}\right)+\red{\mathcal{T}}^-_{j}M_{i}\left(1-\dfrac{M_{j-1}}{m_{j-1}}\right)\nonumber\\
 &&+\red{\mathcal{T}}^+_{j}M_{j}\left(1-\dfrac{M_{j+1}}{m_{j+1}}\right)+\red{\mathcal{T}}^-_{j+1}M_{j+1}\left(1-\dfrac{M_{j}}{m_{j}}\right),
\end{eqnarray}
with equations for the covariances obtainable in a similar manner.

\section{Outline of simulation algorithm}
\label{appendix:SimulationAlgorithm}
In second part of this paper, we compare realisations of four different models of volume-excluding diffusion. The following algorithm was used to generate realisations between $t=0$ and $t=25$:
\begin{enumerate}
 \item Set $timeElapsed=0$ to track the duration of the simulation, and initialise an array to track particle locations.
 \item Generate the exponentially distributed \begin{equation}timeUntilNextEvent=-\mathrm{ln}(rand)/totalEventPropensity,\end{equation} where $rand$ is a uniformly distributed random number in the range $[0, 1]$. 
 \item If $timeElapsed+timeUntilNextEvent>25$, then go to Step 7, otherwise continue to Step 4.
 \item Use a uniformly distributed random number to determine what event has occured. The probability of each event being chosen is given by the array of $eventPropensities$ divided by the $totalEventPropensity$.
 \item Update the array of particle locations to reflect the outcome of the event, and update the individual $eventPropensities$ and the $totalEventPropensity$ accordingly.
 \item Set $timeElapsed=timeElapsed+timeUntilNextEvent$, and return to Step 2.
 \item Store the array describing the particle positions at time $t=25$.
\end{enumerate}
In the first two test cases, the array of $eventPropensities$ consists of left and right jump propensities for each compartment, some of which will be reduced or elimated completely due to filled volumes in neighbouring compartments, and the $totalEventPropensity$ is their sum. For the morphogen gradient case, the $eventPropensities$ will also include decay propensities for each occupied compartment in the system and a constant particle in-flux term. The second test case also records particle positions at $t=1,2,\dots,24$, in addition to the final distribution at $t=25$. For each test case, we generated 50,000 realisations of each model, plus another 50,000 realisations of the fully-excluding model to provide comparison data for the HDE, and then calculated the mean and variance of particle numbers in each compartment. This algorithm follows a deliberately simple design, demonstrating that no advanced programming ability is required to implement any of the models.

\section{Comparison to non-excluding model}

To illustrate the effects of multi-species volume exclusion in Section \ref{sec:SimpleMultiSpeciesExclusionSystem}, we compare the simulation results to the solution of the diffusion equation,
\begin{equation}
 \dfrac{\partial c}{\partial t}=D\dfrac{\partial^2 c}{\partial x^2},
\end{equation}
where $c(x,t)$ is the concentration of particles. This PDE was solved using the Matlab routine \textsf{pdepe}, with grid spacing $1.05\times10^{-2}$ and time step $\Delta t=2.5\times10^{-3}$.

\begin{figure}
\includegraphics[width=18cm]{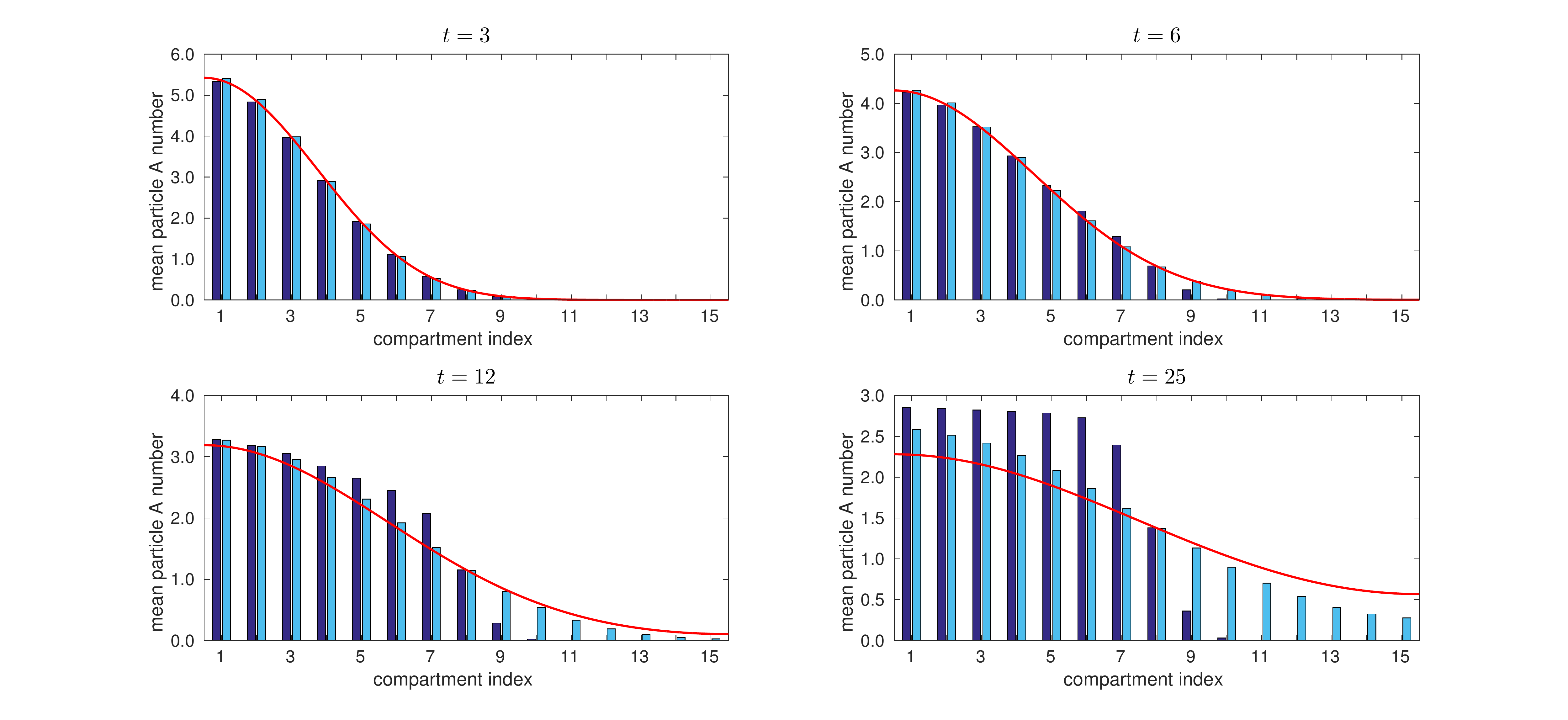}
\caption{A comparison of the results from Section \ref{sec:SimpleMultiSpeciesExclusionSystem} to a non-excluding PDE. The dark and light blue bars represent the mean number of species A particles, while the red line shows the solution of the non-excluding diffusion equation, starting from the same initial conditions. At time $t=3$, when interaction with the species B particles is negligible, all three results agree well; by time $t=6$ the effects of interaction with species B begin to be seen, and the models continue to diverge for the remainder of the simulation.}
\label{fig:FullyAndPartiallyExcluding_CompareWithPDE}
\end{figure}

We plot comparisons between the PDE solution and the mean number of species A particles in Figure \ref{fig:FullyAndPartiallyExcluding_CompareWithPDE}, using both the fully-excluding and partially-excluding models, (similar results can be obtained for species B particles). At time $t=3$, all three sets of results are in agreement as we would expect. There has been little interaction with the species B particles for times $t<3$, so the behaviour of species A particles can be approximated as a single species model. The fully-excluding results therefore match both the partially-excluding results and the diffusion equation. As $t$ increases, however, the effects of volume exclusion become apparent and all three sets of results begin to diverge. As an aside, we note that if the mean total number of particles was plotted, counting both species A and B, then the resulting simulation results would be in agreement with the PDE~\cite{Simpson:2009:MSE}.

\section{Two-dimensional models}

\begin{figure}
\includegraphics[width=12cm]{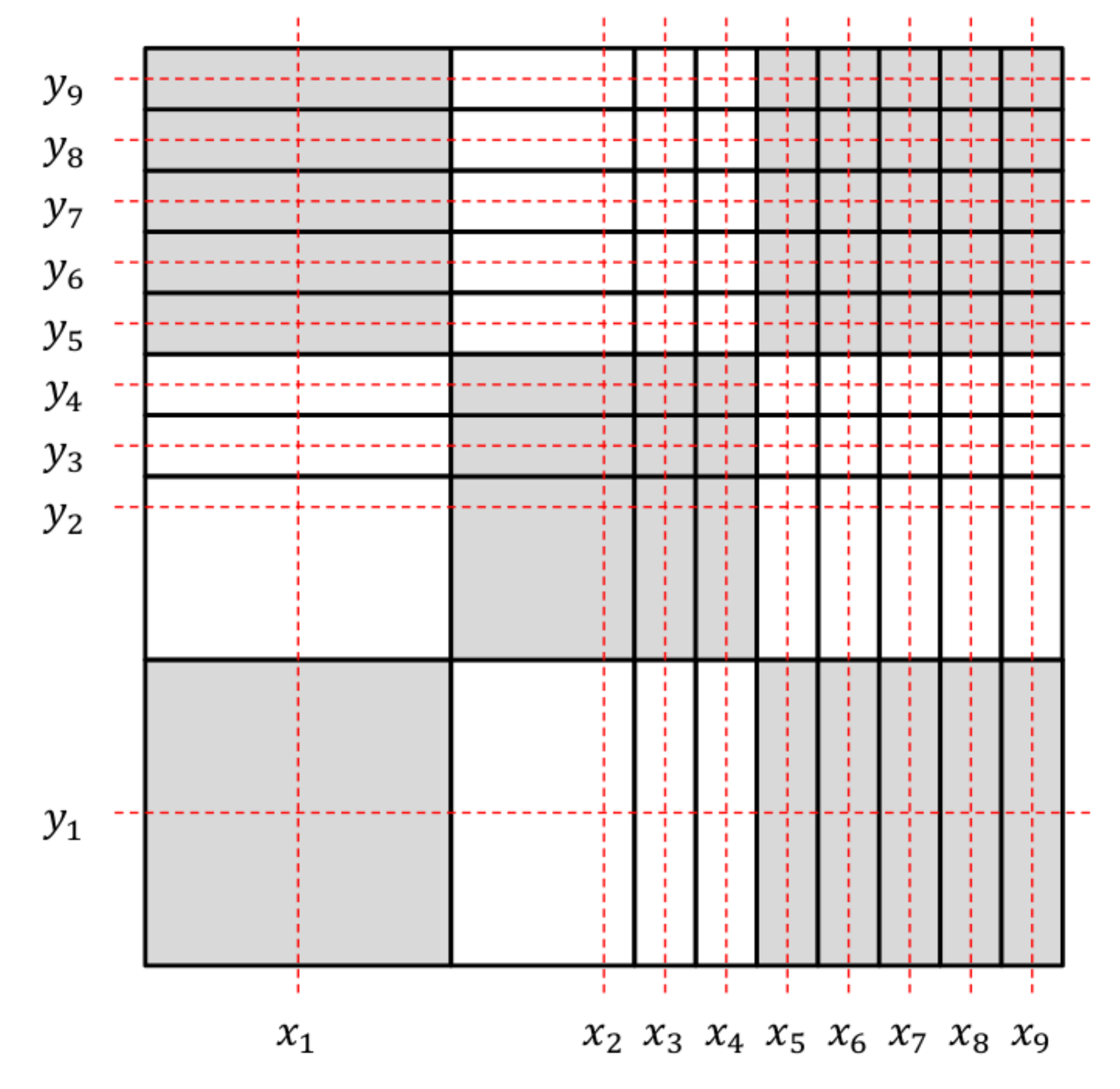}
\caption{An example rectilinear Voronoi partition in two spatial dimensions, illustrating one way in which a coarse-grained region in the bottom-left can be connected to the fine-grained region in the top-right. Residence points are located at the intersections of the red dashed lines and black lines mark compartment boundaries. Grey and white shading is used to indicate regions which would be aggregated to obtain $S_{i,j}^{(m)}$ terms.}
\label{fig:RectilinearVoronoi2d_Cropped}
\end{figure}

In this section, we sketch out how the hybrid methods could be extended to two spatial dimensions. For a rectilinear Voronoi partition, of the kind illustrated in Figure \ref{fig:RectilinearVoronoi2d_Cropped}, we demonstrate that the mean master equation remains linear.
We write $n^{(v)}_{i,j}$ and $m_{i,j}$ to denote the number of particles in, and the capacity of, the compartment with residence point at $(x_i,y_j)$. As before, we write $M^{(v)}_{i,j}=\langle n^{(v)}_{i,j}\rangle$. To denote the distance between residence points, we use $\Delta x_i=x_i-x_{i-1}$ and $\Delta y_j=y_j-y_{j-1}$. By analogy to Eq. (\ref{eq:DefOfCompartmentCapacity}) in the main text, we assume that each particle is of size $h\times h$, and hence write
\begin{equation}
\label{eq:defOfCapacityIn2D} m_{i,j}=\dfrac{(\Delta x_i+\Delta x_{i+1})(\Delta y_j+\Delta y_{j+1})}{4h^{2}},\;\;1<i,j<K,\\
\end{equation}
with similar expressions obtainable for the capacities of compartments at the boundaries of the domain. Transition rates for diffusion on two-dimensional rectilinear Voronoi lattices\red{, in the absence of volume exclusion,} are given in \cite{Yates:2013:IVP} as
\begin{eqnarray}
 \red{\mathcal{T}}^{r}_{i,j}&=&\dfrac{2Dm_{i+1,j}}{m_{i,j}\Delta x_{i+1}(\Delta x_{i+1}+\Delta x_{i+2})},\\
 \red{\mathcal{T}}^{d}_{i,j}&=&\dfrac{2Dm_{i,j-1}}{m_{i,j}\Delta y_{j}(\Delta y_{j-1}+\Delta y_{j})},\\
 \red{\mathcal{T}}^{l}_{i,j}&=&\dfrac{2Dm_{i-1,j}}{m_{i,j}\Delta x_{i}(\Delta x_{i-1}+\Delta x_{i})},\\
 \red{\mathcal{T}}^{u}_{i,j}&=&\dfrac{2Dm_{i,j+1}}{m_{i,j}\Delta y_{j+1}(\Delta y_{j+1}+\Delta y_{j+2})},
\end{eqnarray}
where $\red{\mathcal{T}}^{r}_{i,j},\red{\mathcal{T}}^{d}_{i,j},\red{\mathcal{T}}^{l}_{i,j}$ and $\red{\mathcal{T}}^{u}_{i,j}$ denote, respectively, the transition rates from the compartment indexed $(i,j)$ to the compartments indexed $(i+1,j)$, $(i,j-1)$, $(i-1,j)$ and $(i,j+1)$ (i.e. movement up, right, down and left). \red{After including standard blocking probabilities, we} hence obtain the mean master equation
\begin{eqnarray}
 \label{eq:2dMasterMeanEqWithNonlinTerms}\dfrac{\mathrm{d}M^{(v)}_{i,j}}{\mathrm{d}t}=&&\red{\mathcal{T}}^{l}_{i+1,j}\left(M^{(v)}_{i+1,j}-\dfrac{1}{m_{i,j}}\langle n^{(v)}_{i,j}n^{(v)}_{i+1,j}\rangle\right)-\red{\mathcal{T}}^{r}_{i,j}\left(M^{(v)}_{i,j}-\dfrac{1}{m_{i+1,j}}\langle n^{(v)}_{i,j}n^{(v)}_{i+1,j}\rangle\right)\nonumber\\
 &+&\red{\mathcal{T}}^{u}_{i,j-1}\left(M^{(v)}_{i,j-1}-\dfrac{1}{m_{i,j}}\langle n^{(v)}_{i,j-1}n^{(v)}_{i,j}\rangle\right)-\red{\mathcal{T}}^{d}_{i,j}\left(M^{(v)}_{i,j}-\dfrac{1}{m_{i,j-1}}\langle n^{(v)}_{i,j-1}n^{(v)}_{i,j}\rangle\right)\nonumber\\
 &+&\red{\mathcal{T}}^{r}_{i-1,j}\left(M^{(v)}_{i-1,j}-\dfrac{1}{m_{i,j}}\langle n^{(v)}_{i-1,j}n^{(v)}_{i,j}\rangle\right)-\red{\mathcal{T}}^{l}_{i,j}\left(M^{(v)}_{i,j}-\dfrac{1}{m_{i-1,j}}\langle n^{(v)}_{i-1,j}n^{(v)}_{i,j}\rangle\right)\nonumber\\
 &+&\red{\mathcal{T}}^{d}_{i,j+1}\left(M^{(v)}_{i,j+1}-\dfrac{1}{m_{i,j}}\langle n^{(v)}_{i,j}n^{(v)}_{i,j+1}\rangle\right)-\red{\mathcal{T}}^{u}_{i,j}\left(M^{(v)}_{i,j}-\dfrac{1}{m_{i,j-1}}\langle n^{(v)}_{i,j-1}n^{(v)}_{i,j}\rangle\right),
 \end{eqnarray}
where the four rows describe, respectively, the exchange of particles between compartment $(i,j)$ and the compartments right, below, left and above it. Examining the non-linear terms on the first row, we note that
\begin{eqnarray}
 \left(\dfrac{-\red{\mathcal{T}}^{l}_{i+1,j}}{m_{i,j}}+\dfrac{\red{\mathcal{T}}^{r}_{i,j}}{m_{i+1,j}}\right)\langle n^{(v)}_{i,j}n^{(v)}_{i+1,j}\rangle&=&\left(\dfrac{-2Dm_{i,j}}{m_{i,j}m_{i+1,j}\Delta x_{i+1}(\Delta x_{i}+\Delta x_{i+1})}\right.\nonumber\\
 &&+\left.\dfrac{2Dm_{i+1,j}}{m_{i,j}m_{i+1,j}\Delta x_{i+1}(\Delta x_{i+1}+\Delta x_{i+2})}\right)\langle n^{(v)}_{i,j}n^{(v)}_{i+1,j}\rangle\nonumber\\
 &=&\dfrac{2D}{m_{i,j}m_{i+1,j}\Delta x_{i+1}}\left(\dfrac{-m_{i,j}}{\Delta x_{i}+\Delta x_{i+1}}+\dfrac{m_{i+1,j}}{\Delta x_{i+1}+\Delta x_{i+2}}\right).\nonumber
\end{eqnarray}
Using the definition of capacity from Eq. (\ref{eq:defOfCapacityIn2D}), it can be seen that the terms inside the brackets of the final line cancel out, and hence the contribution from the non-linear terms is zero. Similar reasoning can be applied to the other lines of Eq. (\ref{eq:2dMasterMeanEqWithNonlinTerms}) to obtain
\begin{eqnarray}
 \dfrac{\mathrm{d}M^{(v)}_{i,j}}{\mathrm{d}t}=&&\red{\mathcal{T}}^{l}_{i+1,j}M^{(v)}_{i+1,j}-\red{\mathcal{T}}^{r}_{i,j}M^{(v)}_{i,j}\nonumber\\
 &+&\red{\mathcal{T}}^{u}_{i,j-1}M^{(v)}_{i,j-1}-\red{\mathcal{T}}^{d}_{i,j}M^{(v)}_{i,j}\nonumber\\
 &+&\red{\mathcal{T}}^{r}_{i-1,j}M^{(v)}_{i-1,j}-\red{\mathcal{T}}^{l}_{i,j}M^{(v)}_{i,j}\nonumber\\
 &+&\red{\mathcal{T}}^{d}_{i,j+1}M^{(v)}_{i,j+1}-\red{\mathcal{T}}^{u}_{i,j}M^{(v)}_{i,j},
 \end{eqnarray}
which is linear, as expected from the corresponding mean master equation in one spatial dimension.

\begin{figure}
\includegraphics[width=10cm]{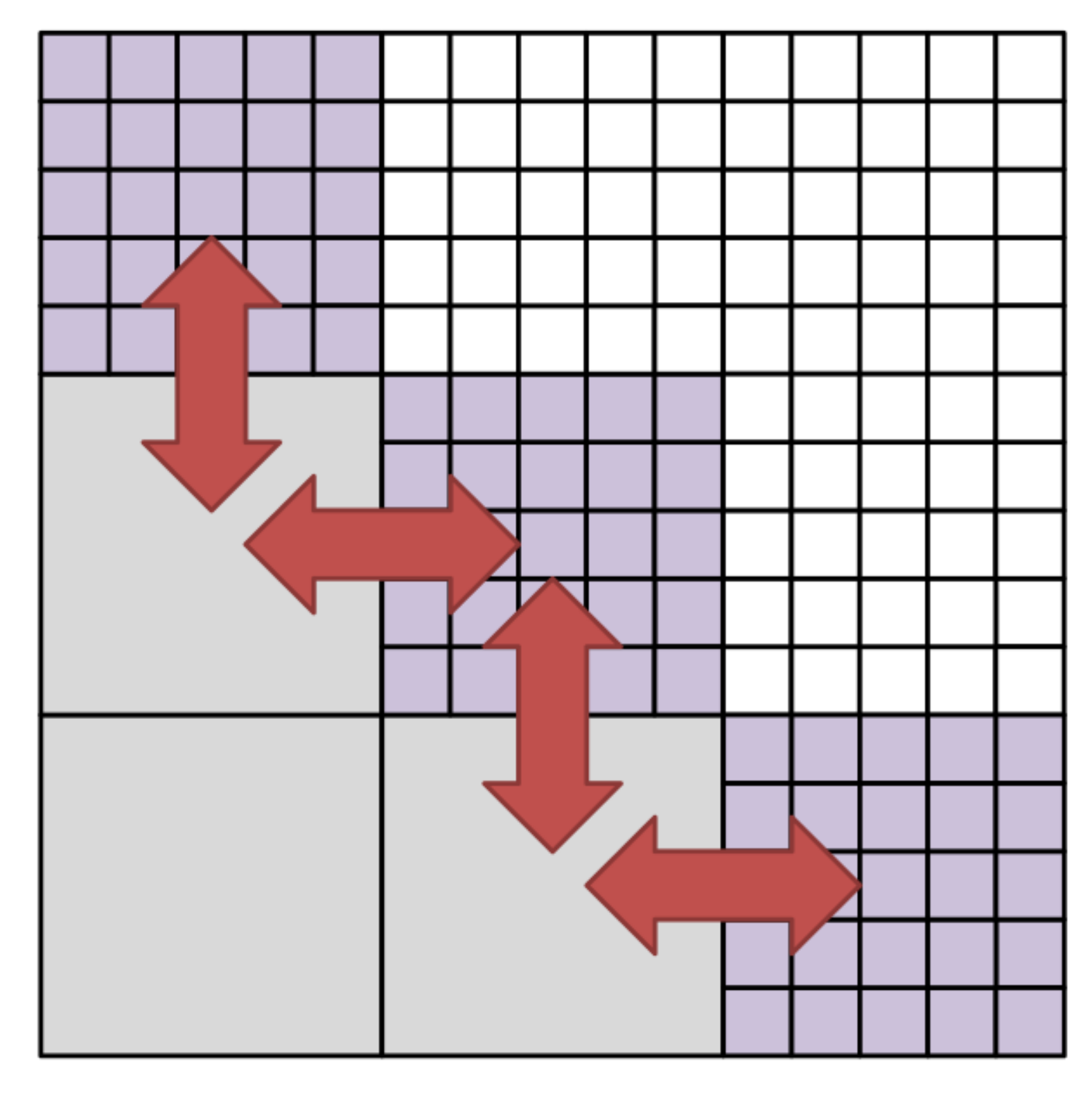}
\caption{An example pseudo-compartment method in two spatial dimensions. Partially-excluding compartments are shown in grey. }
\label{fig:PseudoCompartmentExample2d_Cropped}
\end{figure}

We do not present the details here, but the pseudo-compartment method could also be extended to two or three dimensions, with one illustrative example presented in Figure \ref{fig:PseudoCompartmentExample2d_Cropped}.

\end{document}